\pdfoutput=1
\documentclass[floatfix,twocolumn,preprintnumbers,amsmath,amssymb,superscriptaddress]{revtex4-2}
\usepackage{amssymb}
\usepackage{amsmath}
\usepackage{makecell}
\usepackage{color}
\usepackage[pdftex]{graphicx}
\usepackage{multirow}
\usepackage{float}
\usepackage{mathtools}
\usepackage{comment}
\usepackage{colortbl}
\usepackage{nicematrix}
\usepackage[toc,page]{appendix}
\DeclareUnicodeCharacter{2212}{-}

\usepackage{makecell}
\usepackage{enumerate}
\usepackage{mathtools}
\usepackage{stmaryrd}
\usepackage{enumitem}
\usepackage{textcomp}
\usepackage{gensymb}
\usepackage{float}
\def \equi#1{\mathrel{\mathop{\kern 0pt\sim}\limits_{#1}}} 

\usepackage{diagbox}

\usepackage[overload]{empheq}
\usepackage{cases}

\def\blue{\textcolor{blue}}
\def\blue{}

\begin{document}
\title{Universal exploration dynamics of random walks}

\author{L\'eo R\'egnier}
\address{Laboratoire de Physique Th\'eorique de la Mati\`ere Condens\'ee,
CNRS/Sorbonne University, 4 Place Jussieu, 75005 Paris, France}
\author{Maxim Dolgushev}
\address{Laboratoire de Physique Th\'eorique de la Mati\`ere Condens\'ee,
CNRS/Sorbonne University, 4 Place Jussieu, 75005 Paris, France}
\author{S. Redner}
\address{Santa Fe Institute, 1399 Hyde Park Road, Santa Fe, NM, USA 87501}
\author{Olivier B\'enichou}
\email{benichou@lptmc.jussieu.fr}
\address{Laboratoire de Physique Th\'eorique de la Mati\`ere Condens\'ee,
CNRS/Sorbonne University, 4 Place Jussieu, 75005 Paris, France}

\begin{abstract}
The territory explored by a random walk is a key property that may be quantified by the number of distinct sites that the random walk visits up to a given time. We introduce a more fundamental quantity, the time $\tau_n$ required by a random walk to find a site that it never visited previously when the walk has already visited $n$ distinct sites, which encompasses the full dynamics about the visitation statistics. To study it, we develop a theoretical approach that relies on a mapping with a trapping problem, in which the spatial distribution of traps is continuously updated by the random walk itself.  Despite the geometrical complexity of the territory explored by a random walk, the distribution of the $\tau_n$ can be accounted for by simple analytical expressions. Processes as varied as regular diffusion, anomalous diffusion, and diffusion in disordered media and fractals, fall into the same universality classes.

%LEO'S ABSTRACT%In this work, we are interested in characterizing the exploration statistics of a random walker. To obtain it, we introduce the time $\tau_n$ required by a random walk to find a site that it never visited previously when the walk has already visited $n$ distinct sites. To determine the distribution of these inter-visit times $\tau_n$, we develop a theoretical approach that relies on a mapping with a trapping problem, in which, in contrast to previously studied situations, the spatial distribution of traps is continuously updated by the random walk itself.  Despite the geometrical complexity of the territory explored by a random walk, the distribution of the $\tau_n$ can be accounted for by simple universal analytical expressions for various processes (regular diffusion, anomalous diffusion, and diffusion in disordered media and fractals).  We confirm our theoretical predictions by Monte Carlo and exact enumeration methods, and extend our approach to other exploration observables. \\
\end{abstract}
\maketitle

\begin{figure}[h!]
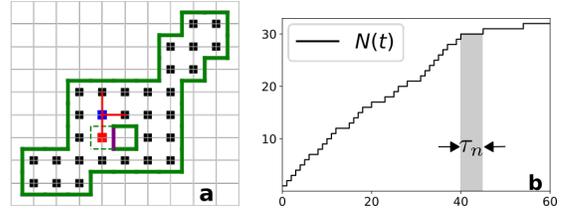

\centering
\includegraphics[width=0.35\columnwidth]{Fig1a.pdf}\quad
\includegraphics[width=0.45\columnwidth]{Fig1b.pdf}
\caption{{\bf a} A visited domain (black sites) and its boundary (green line) for a RW on the square lattice. The $n^{\rm th}$ and $(n+1)^{\rm st}$ new sites visited are blue and red squares. The red links indicate the intervening RW trajectory, and the new boundary when the $(n+1)^{\rm st}$ site is visited is purple. {\bf b} The time intervals $\tau_n$ between increments in $N(t)$, the number of new sites visited at time $t$.} 
    \label{fig:MainIllustration}
\end{figure}

The number $N(t)$ of distinct sites visited by a random walker (RW) up to time $t$ is a key property in random walk theory \cite{hollander_weiss_1994, vineyard1963number,weiss1994aspects,feller2008introduction,Gall_1991, Vallois_1996,klafter2011first} which appears in many physical \cite{scher1980field,shlesinger1984williams,klafter1986relationship,haus1987diffusion,larralde1992territory,shlesinger1992new,song2010modelling,sokolov1997paradoxal,meroz2013test,agliari2007random,burov2012weak,miyazaki2013quantifying,de2014navigability,barkai2015universal}, chemical \cite{rice1985diffusion,havlin1984trapping}, and ecological \cite{gordon1995development} phenomena. This observable quantifies the efficiency of various stochastic exploration processes, such as animal foraging \cite{gordon1995development} or the trapping of diffusing molecules \cite{havlin1984trapping,hollander_weiss_1994}. While the average and, for some examples, the distribution of the number of distinct sites visited, have been determined analytically~\cite{Gall_1991,gillis_2003,biroli2022number,Leo_2022}, this information is far from a complete description. In this work, we show that the waiting time $\tau_n$, defined as the elapsed time between the visit to the $n^{\rm th}$ and the $(n+1)^{\rm st}$ distinct, \blue{or new,} sites characterizes the exploration dynamics in a more fundamental and comprehensive way (Figure~\ref{fig:MainIllustration}). 

\blue{In addition to their basic role in characterizing site visitation, the $\tau_n$ are central to phenomena that are controlled by the time between visits to new sites. A class of such models are self-interacting RW, where a random walker deposits a signal at each visited site that alters the future dynamics of the walker on its next visit to these sites. This self-attracting random wal \cite{barbier2022self,barbier2020anomalous,Sapozhnikov_1994} has recently been shown to account for real trajectories of living cells \cite{Nature_Alex}. In this model, the probability that the RW jumps to a neighboring site $i$ is proportional to $\exp(−u n_i )$, where $n_i = 0$ if the site $i$ has never been visited up to time $t$ and $n_i = 1$ otherwise.} \blue{The analysis of this strongly non-Markovian walk is a difficult problem with few results available in dimension higher than 1. However, we note that its evolution between visits to new sites is described by a regular random walk whose properties are well known. This makes the determination of the statistics of the $\tau_n$ 
an important first step in the analysis and understanding of these non-Markovian RWs.}

\blue{The variables $\tau_n$ also underlie starving RWs \cite{Benichou_Redner_2014,sanhedrai2021,sanhedrai2020lifetime,benichou2016role,krishnan2018optimizing}, which describe depletion-controlled starvation of a RW forager.} \blue{In these models, the RW survives only if the time elapsed until a new food-containing site is visited is less than an intrinsic metabolic time ${\cal S}$. If the forager collects a unit of resource each time a new site is visited, then in one trajectory, the forager might find resources at an almost regular rate while in another trajectory, the forager might find most of its resources near the end of its wandering.  This discrepancy in histories has dramatic effects: the forager survives on the first trajectory but not the latter. To understand this disparity requires knowledge of the random variables $\tau_n$. }

Despite their utility and fundamentality, the statistical properties of the $\tau_n$ appear to be mostly unexplored, except for the one-dimensional ($1d$) nearest-neighbor RW.  In this special case, the distribution of $\tau_n$ coincides with the classic first-exit probability of a RW from an interval of length $n$, $F_n(\tau)$~\cite{redner_2001}. We drop the subscript $n$ on $\tau$ henceforth, because the value of $n$ will be evident by context. In the limit $n\to\infty$ with $\tau/n^2$ fixed, $F_n$ has the following basic properties: (i) \blue{aging \cite{Levernier2018}; in general, $F_n$ depends explicitly on $n$, or equivalently, the time elapsed until the visit to the $n^{\rm th}$ new site}; (ii) an $n$-independent algebraic decay: $\tau^{-3/2}$ for $1\ll\tau\ll n^2$, where $n^2$ is the typical time to diffuse across the interval; (iii) an exponential decay for $\tau\gg n^2$; (iv) $F_n$ admits the scaling form $F_n(\tau)=n^{-3}\psi\left(\tau/n^2\right)$ (see Sec.~S1 in the Supplementary Information (SI) for details).

\begin{figure*}[th!]
\centering
\includegraphics[width=13em]{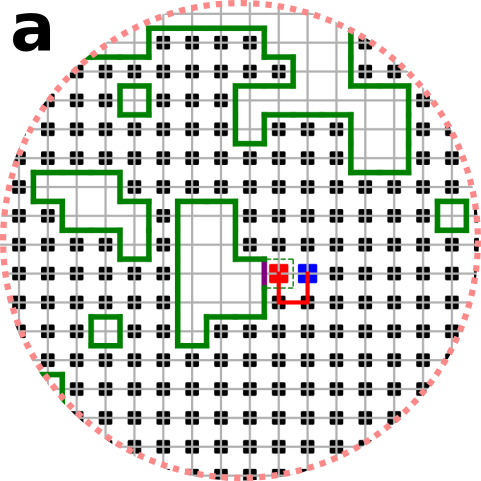}\quad\qquad
\includegraphics[width=20em]{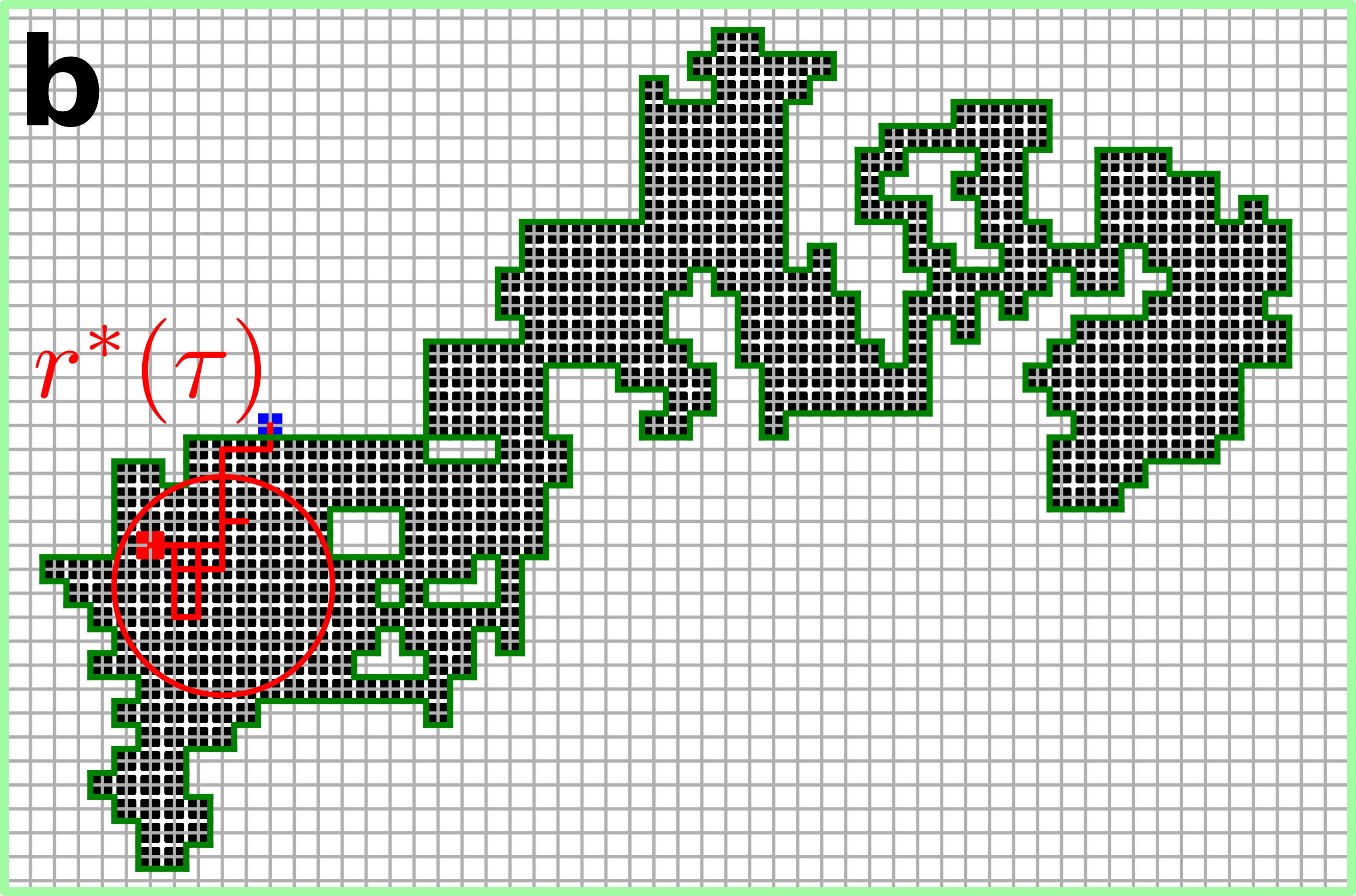} \quad\qquad
\includegraphics[width=13em]{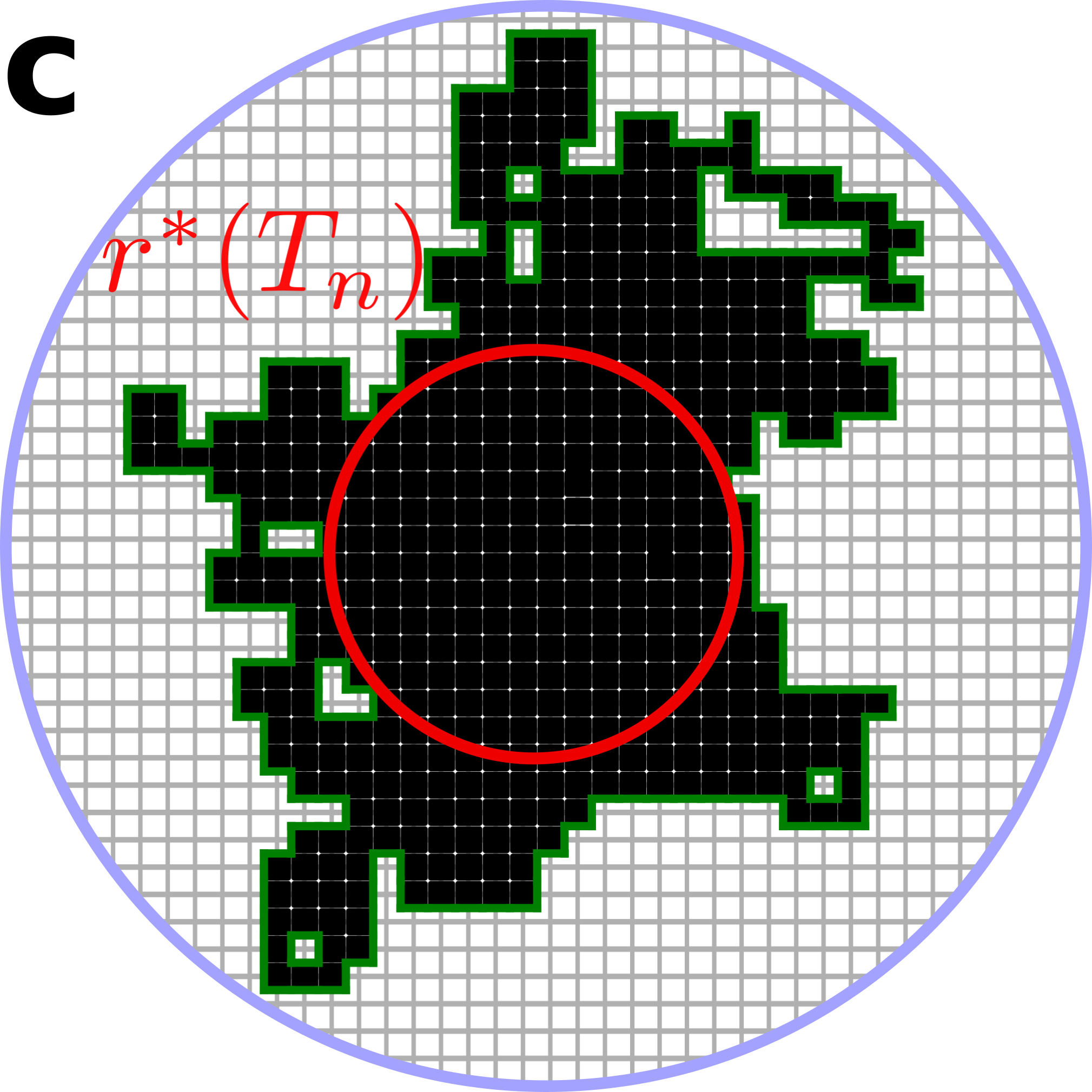}
\caption{\label{fig:cases}The three temporal regimes of the exploration dynamics, as illustrated by a RW on a square lattice.  Each panel shows the corresponding different controlling  configurations when $n=500$ distinct sites have been visited. The $n^{\rm th}$ and $(n+1)^{\rm st}$ visited site are shown in red and blue, respectively ({\bf a} and {\bf b}). \textbf{a} Early time: the visited domain (black squares within the green boundary) is effectively infinite \blue{(at the scale of the trajectory of the RW during the time $\tau_n$)}. \textbf{b} Intermediate time: the exit time probability from the visited domain is governed by atypically large trap-free regions of radius $r^*(\tau) \sim  \rho_n \left( \tau/t_n \right)^{1/(d_\text{f}+d_\text{w})}$. \textbf{c} Long time: the exit time probability is determined by atypically large trap-free regions  of radius $r^*(T_n)\sim n^{1/d_\text{f}}$.}
\end{figure*}

\begin{table*}
\begin{tabular}{c|c|c||@{\hspace{0cm}}c@{\hspace{0cm}}|@{\hspace{0cm}}c@{\hspace{0cm}}|@{\hspace{0cm}}c@{\hspace{0cm}}}
     & $ t_n $ & $T_n$  & \makecell{$1 \ll \tau \ll t_n$ } & \makecell{$t_n \ll \tau \ll T_n$ } & \makecell{$T_n \ll \tau $ }  \\
    \hline
      $\mu<1$ [recurrent]& $n^{1/\mu}$ & $n^{1/\mu}$ & \cellcolor[rgb]{1,0.6,0.6}&\cellcolor[rgb]{0.8,0.8,0.8} & \cellcolor[rgb]{0.6,0.6,1}   \\
     $\mu=1$ [marginal]& $\sqrt{n}$ & $n^{3/2}$\ & \multirow{-2}*{\cellcolor[rgb]{1,0.6,0.6} $\tau^{-1-\mu}\equiv \tau^{-(2-\theta)}$ } & \cellcolor[rgb]{0.6,1,0.6} & \cellcolor[rgb]{0.6,0.6,1} $\exp  \Big[ -\text{const} \; \tau/n^{1/\mu}$ \Big] \\
     $\mu>1$ [transient] & 1 & $n^{1/\mu+1}$ &\cellcolor[rgb]{0.8,0.8,0.8} & \multirow{-2}*{\cellcolor[rgb]{0.6,1,0.6} $\exp \Big[ -\text{const}\left(\tau/t_n \right)^{\frac{\mu}{1+\mu}}$ \Big]} &\cellcolor[rgb]{0.6,0.6,1}  \\
\end{tabular}
\caption{Summary of the time dependence of $F_n(\tau)$ for the three classes of RWs---recurrent, marginal, and transient. The constants are independent of $n$ and $\tau$. The crossover times $t_n$ and $T_n$ are given up to logarithmic prefactors. \blue{The time regimes identified in the last three columns are the same as the ones presented in Figure~\ref{fig:cases}. The persistence exponent $\theta$ is here given by $\theta=1-\mu$, see text.}}
\label{tab:Summary} 
\end{table*}

\section*{Results}

In this work, we extend these visitation properties to the physically relevant and general situations of higher dimensions and general classes of RWs, including anomalous diffusion. We investigate symmetric Markovian RWs that move in a medium of fractal dimension $d_\text{f}$, and whose mean-square displacement is assumed to be given by $\langle {\bf r}^2(t)\rangle \propto t^{2/d_\text{w}}$, where $d_\text{w}$ is the dimension of the walk \cite{ben_Avraham} and $t$ the number of RW steps. We assume in particular the existence of a renewal equation between the propagator and the first-passage time density of the RW \cite{Hughes_1995}. We focus on discrete time and space RWs, for which the number of sites visited at a certain time is clearly defined (see SI S5.C.{\it1} for the extension of our results to  Continuous Time Random Walks (CTRWs)). 
The ratio $\mu={d_\text{f}}/{d_{\text{w}}}$ determines whether the RW is recurrent ($\mu<1$), marginal ($\mu=1$), or transient ($\mu>1$). For recurrent and marginal RWs ($\mu \leq 1$), the probability to eventually visit any site is one, while for transient RWs ($\mu>1$), the probability to visit any site is strictly less than one~\cite{feller1,morters2010brownian}. Despite the geometrical complexity of the territory explored after $n$ steps (which typically contains holes, islands \cite{caser1996} and is not spherical \cite{rudnick1987shapes,rudnick1991shapes}, see Figs.~\ref{fig:MainIllustration} and \ref{fig:cases}), the distribution of the times $\tau$ between visits to new sites obeys universal statistics that are characterized only by $\mu$, as summarized in Table~\ref{tab:Summary} (up to constant prefactors that are independent of $\tau$, and neglecting algebraic corrections for the two latter regimes).  

Fundamental consequences of our results include the following: (i) Finding new sites takes progressively more time for recurrent and marginal RWs; this agrees with simple intuition. This property is quantified by the $n$ dependence of the moments of $\tau_n$.  From the entries in Table~\ref{tab:Summary} we find $\langle \tau_n^k\rangle \propto n^{{k}/{\mu}-1}$ for recurrent RWs, while $\langle \tau_n\rangle \propto \ln n$ and $\langle \tau_n^k\rangle \propto n^{(k-1)/2}$ for $k > 1$ for marginal RWs. Conversely, transient random walks rarely return to previously visited sites, so that $\langle \tau_n^k\rangle \propto {\rm const}$ (see the SI Sec.~S3.D for the derivation \blue{and numerical check}). (ii) The statistics of the $\tau_n$ exhibit universal and giant fluctuations for recurrent and marginal RWs, 
with ${\rm Var}(\tau_n)/\langle \tau_n\rangle^2\propto n$ for recurrent walks and ${\rm Var}(\tau_n)/\langle \tau_n\rangle^2\propto \sqrt{n}/(\ln n)^2$ for marginal walks. In the context of the foraging process mentioned above, this leads to very different life histories of individual foragers. In contrast, $\tau_n$ remains bounded for large $n$ for transient RWs, so that fluctuations remain small. (iii) The early-time regime is independent of $n$.  The feature of aging, which originates from the finite size $n$ of the domain visited, arises after a time $t_n$, for recurrent and marginal RWs, and $T_n$, for transient RWs (see Table ~\ref{tab:Summary} and below for the definition of these two fundamental time scales). (iv) As shown below, each regime of the exploration dynamics is controlled by specific configurations that are illustrated in Figure~\ref{fig:cases}. These provide the physical mechanisms that underlie the entries in Table~\ref{tab:Summary}. \blue{(v) The algebraic decay of $F_n(\tau)$ in the recurrent case should be compared with the simpler problem of a recurrent RW in unbounded space, where the ﬁrst-passage time distribution to a given target
behaves at large times like $F_{\rm target}(\tau)\propto 1/\tau^{1+\theta}$, with $\theta$ the so-called persistence exponent \cite{Schehr_2013}. Because  $\theta=1-\mu$ for processes with stationary increments \cite{Levernier2018}, and in particular for Markovian processes, the algebraic decay of $F_n$ in Table~\ref{tab:Summary} can be rewritten as $F_n(\tau)\propto \tau^{-(2-\theta)}$, in sharp contrast with the decay of $F_{\rm target}(\tau)$. While the two exponents  coincide for a simple random walk in $1d$ (for which $\theta=1/2$), the problem here involves the first-exit time statistics from a domain whose complex shape is generated by the RW itself. } 

We now sketch how to derive these results (see Secs.~S2--S3 of the SI for detailed calculations). As an essential step, we first map the visitation problem to an equivalent trapping problem. In our visitation problem, we view unvisited sites as traps for the RW, so that a RW is trapped whenever it leaves the domain of already visited sites. Here, the term trapped does not mean that the RW disappears, but rather, the RW continues its motion but now with the visited domain expanded by the site just visited and the inter-visit time $\tau$ is reset to zero. By this equivalence to trapping, the time $\tau$ between visits to the $n^{\rm th}$ and $(n+1)^{\rm st}$ new sites is the same as the probability for the RW to first exit the domain that is comprised of the $n$ already visited sites, or equivalently the domain free of traps. A crucial feature of this equivalence to trapping is that the spatial distribution of traps is continuously updated by the RW trajectory itself. In contrast to the classical trapping problem~\cite{donsker1979number,Grassberger1982}, where permanent traps are randomly distributed, here the spatial distribution of traps \emph{ages} because it depends on $n$. Moreover, successive traps are spatially correlated, with correlations generated by the RW trajectory. 

These two key points are accounted for by the distribution $Q_n(r)$ of the radius of the largest spherical region that is free of traps after $n$ sites have been visited. We show in Sec.~S2.D of the SI that this distribution assumes the scaling form $Q_n(r) \simeq \rho_n^{-1}\,\exp\!\big[\!-a\left(r/\rho_n\right)^{d_\text{f}}\big]$, where $a$ is independent of $n$ and $r$ and the characteristic length $\rho_n$ provides the typical scale of this radius $r$.  Furthermore, the $n$ dependence of  $\rho_n$, which quantifies both aging and correlations between traps, is determined by whether the exploration is recurrent or transient. Specifically, we find $\rho_n=n^{1/d_\text{f}}$ for $\mu<1$, $\rho_n=n^{1/2d_\text{f}}$ for $\mu=1$ and $\rho_n$ of the order of one, up to logarithmic corrections for $\mu>1$ (see Sec.~S2 in the SI). A striking feature of these behaviors is that the exponent changes \emph{discontinuously} when $\mu$ passes through 1.

The corresponding time scales $t_n=\rho_n^{d_\text{w}}$ and $T_n$ delineate the three regimes of scaling behaviors summarized in Table~\ref{tab:Summary} and Figure~\ref{fig:cases}: (i) a short-time algebraic regime ($1\ll \tau \ll t_n$), (ii) an intermediate-time stretched exponential regime ($t_n\ll \tau \ll T_n$), and (iii) a long-time exponential regime ($T_n\ll \tau$). Here $T_n$ is defined as the time at which the radius of the trap-free region $r^*(\tau)$ that controls the dynamics takes its maximal possible value of $r_\text{max}=n^{1/d_f}$ (see Figure~\ref{fig:cases}{\textbf c} and the discussion below Eq.~\eqref{bound-2nd}). We do not characterize the early time regime $\tau=O(1)$ which depends on details of the model. We are only interested in universal features.

\smallskip
\emph{Algebraic regime}: Here, the distribution of $\tau$ has a universal algebraic decay whose origin stems from two essential features: (i) The RW just visited a new site so that the RW  starts from the interface between traps and visited sites when the clock for the next $\tau$ begins. (ii) The region already visited by the RW is sufficiently large so that we can treat the region as effectively infinite (Figure~\ref{fig:cases}{\bf a}) and thereby approximate $F_n(\tau)$ by $F_{\infty}(\tau)$.

The first-return time distribution to this set of traps on the interface is determined by the renewal equation~\cite{Hughes_1995,noh2004random,Plyukhin2016} \blue{that links the probability $P_{\rm trap}(t)$ to be at a trap at time $t$ and the distribution of first arrival times $F_\infty(\tau)$ to a trap at time $\tau$,}
\begin{equation}
\label{renewal}
    P_{\rm trap}(t)=\delta(t)+\int_0^{t} F_\infty(\tau)\;P_{\rm trap}(t-\tau)\,d\tau\,.
\end{equation}
This equation expresses the partitioning of the total RW path to the interface into a first-passage path to the interface over a time $\tau$ and a return path to the interface over the remaining time $t-\tau$; here we use a continuous-time formulation for simplicity. In this mean-field type equation (detailed in SI Sec.~S3.A.{\it1} and {\it2}, and supported by numerical simulations given below and an alternative derivation for the exponent of the algebraic decay given in Sec.~S3.A.{\it3} in the SI), we treat the set of traps collectively, which amounts to neglecting correlations between the return time and the location of the traps on the interface.

Next, we estimate $P_{\rm trap}(t)$ by using  the fact that the RW is almost uniformly distributed in  a sphere of radius $r(t)\propto t^{1/d_\text{w}}$ at time $t$. The number of traps within this sphere is given by $r(t)^{d_\text{T}}$. Here $d_\text{T}$ is the fractal dimension of the interface between visited and non-visited sites; as shown in the SI Sec.~S3.A.{\it2}, $d_\text{T}=2d_\text{f}-d_\text{w}$ \blue{for the recurrent case}. Finally, we obtain the fraction of traps within this sphere and thereby $P_{\rm trap}(t)$:
\begin{equation}
\label{estimate_Ptrap}
    P_{\rm trap}(t)\propto \frac{\text{Number of traps}}{\text{Number of sites}} \propto \frac{r(t)^{d_\text{T}}}{r(t)^{d_\text{f}}}\propto  t^{\mu-1}.
\end{equation}
Based on \eqref{estimate_Ptrap}, we solve Eq.~\eqref{renewal} in the Laplace domain and invert this solution to obtain the algebraic decay $F_\infty(\tau)\simeq A\tau^{-1-\mu}$ in Table~\ref{tab:Summary} in the early-time regime for recurrent and marginal RWs (this derivation is given in Sec.~S3.A in the SI, including exact and approximate expressions for the amplitude $A$ for marginal and recurrent RWs, respectively).

In the transient case, the RW is always close to a non-visited site by the very nature of  transience. Consequently, the time scale $t_n$ is of order one and the algebraic regime does not exist.

\smallskip
\emph{Intermediate- and long-time regimes}. If the RW survives beyond the early-time regime, it can now be considered to start from within the interior of the domain of visited sites. In analogy with the classical trapping problem, a lower bound for the survival probability of the RW,  $S_n(\tau)$, is just the probability for the RW to remain within this domain.  This lower bound is controlled by the rare configurations of large spherical trap-free regions in which the RW starts at the center of this sphere, whose radius distribution $Q_n(r)$ was given above.

We develop a large-deviation approach, in which this lower bound is given by the probability $q_n$ for the RW to first survive up to the first crossover time $t_n$, multiplied by the probability for the RW to remain inside a spherical trap-free domain over a time $\tau$. The quantity $q_n$ is given by $\int_{t_n}^\infty F_\infty(\tau) {\rm d} \tau$, which scales as $1/t_n ^{\mu}$ if $\mu\le 1$, and is of order one if $\mu>1$. The probability for the RW to remain inside a spherical domain of radius $r$ over a time $\tau$ asymptotically scales as $\exp(-b\,\tau/r^{d_\text{w}})$, where $b$ is a constant \cite{ben_Avraham}.  As stated above, the probability to find a spherical trap-free region of radius $r$ is given by $Q_n(r)\simeq\rho_n^{-1}\exp[-a(r/\rho_n)^{d_\text{f}}]$.
%, where $a$ is a another constant. 
Summing over all radii up to the largest possible value $r_{\rm max}=n^{1/d_\text{f}}$, we obtain the lower bound 
\begin{eqnarray}
\label{bound-1st}
S_n(\tau)\ge \frac{q_n}{\rho_n} \int_0^{n^{1/d_\text{f}}}\!\!\! \exp\left[-b \tau/r^{d_\text{w}}-a (r/\rho_n)^{d_\text{f}}\right] {\rm d}r\; ,
\end{eqnarray}
where $a$ and $b$ are constants.  Using Laplace's method \blue{ by making the change of variable $r =\rho \tau^{1/(d_\text{f}+d_\text{w})}$, we obtain (ignoring algebraic prefactors in $n$ and $\tau$)},
\begin{eqnarray}
\label{bound-2nd}
&S_n(\tau)\ge \int_0^{n^{1/d_\text{f}}}\!\!\! \exp\left[-\tau^{\mu/(1+\mu)} \left( b/\rho^{d_\text{w}}+a (\rho/\rho_n)^{d_\text{f}}\right)\right] {\rm d}\rho \nonumber \\
&\gtrsim \exp\left[-\tau^{\mu/(1+\mu)} \left( b/\rho^{*d_\text{w}}+a (\rho^*/\rho_n)^{d_\text{f}}\right)\right] \; ,
\end{eqnarray}
where the minimum of $b/\rho^{d_\text{w}}+a (\rho/\rho_n)^{d_\text{f}}$ is reached at the value $\rho^*$. 
The lower bound \eqref{bound-2nd} for $\tau\gg 1$ is controlled by trap-free regions of radius $r^*(\tau)=\rho^* \tau^{1/(d_\text{f}+d_\text{w})} \sim \rho_n^{{\rm d_f}/({\rm d_f}+{\rm d_w})} \tau^{1/(d_\text{f}+d_\text{w})}$ (see SI Sec.~S3.B for details). Using $t_n=\rho_n^{d_\text{w}}$, this optimal radius is then $r^*(\tau)\sim \rho_n (\tau/t_n)^{1/(d_\text{f}+d_\text{w})}$.  For $\tau\gg t_n$, we have $r^*(\tau)\gg \rho_n$.  Since $\rho_n$ determines the typical radius of the largest spherical region free of traps, the configurations that control the long-time dynamics (as illustrated in Figure~2{\bf b} and {\bf c}) are atypically large, and become more so as $\tau$ increases. Thus the survival probability in this long-time regime is determined by a compromise between the scarceness of large trap-free domains and the long exit times from such domains. Finally, we obtain $F_n(\tau)=-{\rm d}S_n(\tau)/{\rm d}\tau\sim \exp \big[-\text{const}  \left(\tau/t_n \right)^{{\mu}/({1+\mu})}\big]$. As in the classic trapping problem \cite{hollander_weiss_1994,weiss1994aspects,Hughes_1995}, we expect that this lower bound for the survival probability will have the same time dependence as the survival probability itself.

This stretched exponential decay holds as long as the optimal radius is smaller than the maximal value $r_{\rm max}$. The point at which this inequality no longer holds defines a second crossover time $T_n$ by $r^*(T_n)=n^{1/{\rm d_f}}$.  Beyond this time, the evaluation of the integral in Eq.~\eqref{bound-2nd} now leads to an exponential decay of $F_n$ (Table~\ref{tab:Summary}). 

\begin{figure*}[th!]
\centering
\includegraphics[width=2\columnwidth]{FigAll.pdf}
\caption{
{\bf Universal distribution of the time between visits to new sites for RWs}.\\
{\bf Recurrent random walks ($\mu<1)$}. Shown is the scaled distribution $Y\equiv \theta_n^{1+\mu}F_n(\tau)$ versus $X\equiv\tau/\theta_n$ for $n=100,\;500$, and $1000$. Here $\theta_n\sim n^{1/\mu}$ is the decay rate of the exponential in $F_n(\tau)\sim \exp\left(-\tau/\theta_n\right)$. The red dashed lines indicate the algebraic decay $A(\mu)t^{-1-\mu}$ ($A(\mu)$ defined in SI Sec.~S3.A.{\it4}). {\bf a} $1d$ L\'evy flights with index $\alpha=1/\mu=\ln 6/\ln 3$. {\bf b} Subdiffusion on a Sierpinski gasket ($\mu=\ln 3/\ln 5$, scaling of $\theta_n$ with $n$ shown in SI Sec.~S4.A). {\bf c} Subdiffusion on a $2d$ critical percolation cluster ($\mu\approx 0.659$).\\
{\bf Marginally recurrent RWs ($\mu=1$)}. {\bf d} {\bf e} Marginal RWs $(\mu=1)$ at early times. Shown is the scaled distribution $Y\equiv nF_n(\tau)$ versus $X\equiv \tau/\sqrt{n}$ for {\bf d} $1d$ L\'evy flights of index $\alpha=1$ for $n=800$, $1600$, and $3200$,  {\bf e} persistent RWs in $2d$ where the probability to continue in the same direction is $p=0.3$ for $n=800$, $1600$ and $3200$. The red dashed line represent the algebraic decay $A\tau^{-2}$ ($A$ given in SI Sec.~S3.A.{\it4}). {\bf f} Marginal RWs at intermediate and long times. Shown is the scaled distribution $Y \equiv \left( -\ln nF_n(\tau)\right)/\sqrt{\tau/\sqrt{n}}$ versus $X\equiv\tau/n^{3/2}$ for simple RWs on a $2d$ square lattice for $n=200$, $800$ and $3200$. The green and blue dashed lines represent the stretched exponential and the exponential regimes, respectively. \\
{\bf Transient RWs $(\mu>1)$}. Shown is the scaled distribution $Y\equiv \left(-\ln F_n(\tau)\right)/\tau^{\mu/(1+\mu)}$ for {\bf g} L\'evy flights of parameter $\alpha=1$ in $2d$, for $n=400$, $800$, $1600$ and $X\equiv \tau$, {\bf h} persistent RWs in $3d$ where the probability to continue in the same direction is $p=0.25$ for $n=200$, $800$, $3200$ and $X\equiv \tau$, {\bf i} simple RWs on cubic lattice, for $n=200$, $400$, $500$ and  $X\equiv\tau/n^{1+1/\mu}$. The green and blue dashed lines represent the stretched exponential and the exponential regimes, respectively.\\
For all panels,  blue stars, orange circles and green squares correspond to increasing values of $n$. The insets indicate the jump processes. Red squares are the initial and arriving positions of the walker. The green squares represent the prior position of the walker.}
\label{fig:All}
\end{figure*}

Finally, note that the full time dependence of $F_n(\tau)$ has a particularly simple form for recurrent RWs. In this case, the intermediate stretched exponential regime does not exist because $t_n$ and $T_n$ both have the same $n$ dependence. In fact, the short- and long-time limits of $F_n(\tau)$ can be synthesized into the scaling form (as  explained in Sec.~S3.C of the SI)
\begin{equation}
\label{scalingcompact}
    F_n(\tau)=\frac{1}{n^{1+1/\mu}} \;\psi \left( \frac{\tau}{n^{1/\mu}}\right ),
\end{equation}
with $\psi$ a scaling function.

 We confirm the validity of our analytical results by comparing them to numerical simulations of paradigmatic examples of RWs that embody the different cases in Table~\ref{tab:Summary}. The recurrent case ($\mu<1$) is illustrated in Figure~\ref{fig:All}{\bf a}, {\bf b} and {\bf c} for diverse processes: superdiffusive L\'evy flights in $1d$, in which the distribution of jump lengths is fat-tailed, $p(\ell)\propto \ell^{-1-\alpha}$, with $\alpha\in ]1,2[$; subdiffusive RWs on deterministic fractals with and without loops, respectively represented by the Sierpinski gasket and the T-tree (see Sec.~S4.A of SI for the definition of the T-tree and the simulation results); subdiffusive RWs on disordered systems, as represented by a critical percolation cluster on a square lattice.  Our simulations confirm the scaling form of $F_n(\tau)$ given in Eq.~\eqref{scalingcompact}, as well as its algebraic ($X\equiv \tau/t_n<1$) and exponential ($X>1$) decays  at respectively short and long times.

The marginal case ($\mu=1$) is illustrated by $1d$ L\'evy flights of parameter $\alpha=1$, persistent and simple RWs on the $2$-dimensional square lattice (Figs.~\ref{fig:All}{\bf d}, {\bf e} and {\bf f} respectively). The data collapse when plotted versus the scaling variable $\tau/\sqrt{n}$; this confirms that the crossover time $t_n$ scales as $t_n\propto\sqrt{n}$. Figures~\ref{fig:All}{\bf d} and {\bf e} clearly show the expected algebraic decay $\tau^{-2}$ at short times (dashed line). Figure~\ref{fig:All}{\bf f} validates the stretched exponential form of $F_n(\tau)$ at intermediate times, as well as the exponential decrease at long times \blue{and the scaling of $T_n=n^{3/2}$}.

The transient case ($\mu>1$) is illustrated by RWs on hypercubic lattices (see Figure~\ref{fig:All}{\bf g} for the $2d$ L\'evy flights of parameter $\alpha=1$ , \blue{Figure~\ref{fig:All}{\bf h} for a persistent RW and Figure~\ref{fig:All}{\bf i} for a nearest neighbour RW in $3d$, as well as Sec.~S4.C.{\it5} in the SI for higher dimensions and Sec.~S4.C.{\it 6} for transient L\'evy flights)}. 
Figure~\ref{fig:All}{\bf i} confirms the stretched exponential temporal decay for intermediate times, the scaling of the crossover time $T_n=n^{1/\mu+1}$, and the long-time exponential decay of $F_n(\tau)$ for transient RWs.  The numerically challenging task of observing the stretched exponential decay followed by the exponential decay that originates from rare, trap-free regions,
was achieved by relying on Monte Carlo simulations coupled with an exact enumeration technique (see Sec.~S4.C of the SI for details). \blue{We note that in Figure~\ref{fig:All}{\bf g} and {\bf h}, the distribution is independent of $n$ for the values of $X\equiv \tau$ represented, and $Y=-\left(\ln F_n(\tau)\right)/\tau^{\mu/(1+\mu)}$ reaches a plateau. It further confirms the stretched exponential regime and the absence of the algebraic regime ($t_n=1$).}

Overall, we find excellent agreement between our analytical predictions and numerical simulations. The diverse nature of these examples also demonstrates the wide range of applicability of our theoretical approach.

We can extend our approach to treat the dynamics of other basic observables that characterize the support of RWs.  Following~\cite{van1996topology,Mariz2001} two classes of observables can be defined: boundary and bulk. Boundary observables involve both visited and unvisited sites, such as the perimeter $P(t)$ of the visited domain or the number of islands $I(t)$ enclosed in the support of the RW trajectory; note that these variables can both increase and decrease with time. We show, for example, in Sec.~S5.A of the SI, that the corresponding distribution of the times between successive increases in a boundary observable $\Sigma$ again has an early-time algebraic decay, $F_\Sigma(\tau)\propto \tau^{-2\mu}$ for $\mu<1$, and $F_\Sigma(\tau)\propto \ln\tau/\tau^2$ for $\mu=1$. These behaviors are illustrated in Figure~\ref{fig:Islands}{\bf a}, {\bf b} and {\bf c}. Bulk observables involve only visited sites, such as the number of dimers~\cite{van1996topology}, $k$-mers, and $k\times k$ squares in $2d$. We show in Sec.~S5.A of the SI that the dynamics of bulk variables is the same as that for the number of distinct sites visited.

\section*{Discussion}

\begin{figure*}[th!]
\centering
%\includegraphics[width=2\columnwidth]{Fig3_2.pdf}
%\caption{{\bf Boundary observables for recurrent and marginal RWs:} the perimeter of the visited domain and the number of islands enclosed in the support. {\bf a} The elapsed time $\tau_P$ for successive increments of the time dependence of the perimeter $P(t)$ of the visited domain. {\bf b} Distribution $F_P(\tau)$ of the time elapsed $\tau_S$ between the first observations of a domain perimeter of length $P$ and $P+2$ for a RW on the square lattice. {\bf c} Distribution $F_I(\tau)$ of the elapsed time $\tau_I$ between the first occurrence of $I$ and $I+1$ islands for a L\'evy flight with index $\alpha=1.2$.  Plotted in {\bf b} and {\bf c} are the scaled distributions $Y\equiv F_P(\tau)/\left(\ln 8\tau\right)$ and  $Y\equiv F_I(\tau)$ versus $X=\tau$.  The red dashed lines have slope $-2\mu$.  The data are for $P, I=50$, 100, and 200 (respectively blue stars, orange circles and green squares). }
\includegraphics[width=2\columnwidth]{FigApplications.pdf}
\caption{
\blue{{\bf Extensions and applications of the time between visits to new sites for RWs.}\\
{\bf Boundary observables for recurrent and marginal RWs:} The perimeter of the visited domain and the number of islands enclosed in the support. {\bf a} The elapsed time $\tau_P$ for successive increments of the time dependence of the perimeter $P(t)$ of the visited domain. {\bf b} Distribution $F_P(\tau)$ of the time elapsed $\tau_P$ between the first observations of a domain perimeter of length $P$ and $P+2$ for simple RWs on the square lattice. {\bf c} Distribution $F_I(\tau)$ of the elapsed time $\tau_I$ between the first occurrence of $I$ and $I+1$ islands for L\'evy flights of index $\alpha=1.2$.  Plotted in {\bf b} and {\bf c} are the scaled distributions $Y\equiv F_P(\tau)/\left(\ln 8\tau\right)$ and  $Y\equiv F_I(\tau)$ versus $X=\tau$.  The red dashed lines have slope $-2\mu$.  The data are for $P, I=50$, 100, and 200 (respectively blue stars, orange circles and green squares).\\
{\bf  Multiple-time covariances and starving RWs.} {\bf d} $Y=\text{Cov}\left[N(t_1),N(t_2) \right]/\left(\left\langle N(t_1) \right\rangle \left\langle N(t_2) \right\rangle \right)$ for L\'evy flights of parameter $\alpha=1.5$, and we compare $Y$ to $\frac{t_1}{t_2}$ (dashed line).  The stars, circles, and squares indicate data for $t_1=10$, $t_1=100$, and $t_1=1000$. {\bf e} $Y=\frac{t_3}{t_1}\left\langle (N(t_1)-\left\langle N(t_1)\right\rangle )(N(t_2)-\left\langle N(t_2)\right\rangle )(N(t_3)-\left\langle N(t_3)\right\rangle )(N(t_4)-\left\langle N(t_4)\right\rangle )\right\rangle /(\left\langle N(t_1)\right\rangle \left\langle N(t_2)\right\rangle\left\langle N(t_3)\right\rangle \left\langle N(t_4)\right\rangle)$, for L\'evy flights of parameter $\alpha=1.3$. We compare $Y$ to the dashed line proportional to $\frac{t_3}{t_4}$. Data in red and green indicate $t_1=10$ and $t_1=100$. Stars indicate $t_2=2t_1$ and circles indicate $t_2=4t_1$. We take $t_3=4t_2$. {\bf f} Lifetime at starvation. Blue circles show the mean lifetime versus the metabolic time $\mathcal{S}$. The dashed line is proportional to $\mathcal{S}^2$.\\
{\bf Non-Markovian examples.} Rescaled distribution $Y=F_n(\tau)n^{1+1/\mu}$ versus $X=\tau/n^{1/\mu}$ for {\bf g} Fractional Brownian motion with parameter $1/H=1/0.4=\mu=1-\theta$ ($n=20$, $40$ and $80$) {\bf h} Fractional Brownian motion with parameter  $1/H=1/0.75=\mu=1-\theta$ ($n=20$, $40$ and $80$), {\bf i} True Self Avoiding Walks $\mu=2/3=1-\theta$ ($n=200$, $400$ and $800$).\\
For the last three panels, increasing values of $n$ are represented successively by blue stars, orange circles and green squares, and the dashed line is proportional to $X^{-(2-\theta)}$.}}
 \label{fig:Islands}
\end{figure*} 

\blue{In addition to providing asymptotic expressions for the $\tau_n$ distribution and their extension to basic observables characterizing the support of RWs, our results open new  avenues in several directions.} \blue{First, they allow us to revisit the old question of the number $N(t)$ of distinct sites visited at time $t$. Indeed,  our theoretical approach for the set of inter-visit times $\tau$  represents a start towards determining multiple-time visitation correlations for general RWs, quantities that have remained inaccessible this far. These multiple-time correlations are crucial to fully characterize the stochastic process $\{N(t)\}$, the number of sites visited at every single time. However, they have been studied only for the special case of $1d$ nearest-neighbor RWs %\sid{I commented out a short phrase here that I think is not useful}%, which have no holes in their trajectories
~\cite{Annesi:2019,Leo_2022}.  Using our formalism we can further compute temporal correlations of $\{N(t)\}$ for compact L\'evy flights in $1d$ with $1/\mu=\alpha>1$ (which \emph{do} leave holes in their trajectories). We compute the scaling with time of the two-time covariance of the number of distinct sites visited,
\begin{align*}
  \text{Cov}[N(t_1),N(t_2)]\equiv\langle N(t_1) N(t_2)\rangle -\langle N(t_1) \rangle \langle  N(t_2)\rangle  \, .
\end{align*} 
}
\blue{We obtain in the limit $1 \ll t_1 \ll t_2$ (see Sec.~S5.B of the SI for a numerical check of the derivation of Eq.~\eqref{cov} and its numerical confirmation which can also be seen in Figure~\ref{fig:Islands}{\bf d}),
\begin{align}
\label{cov}
    \text{Cov}[N(t_1),N(t_2)] \propto  \;t_1^{\mu}\;t_2^{\mu} \;\;\frac{t_1}{t_2} \; .
\end{align}
 This result can be further extended to $k$-time correlation functions (see the numerical confirmation for $k=4$ in Figure~\ref{fig:Islands}{\bf e}), 
    \begin{align}
    &\left\langle  (N(t_1)-\left\langle N(t_1) \right\rangle )\ldots (N(t_k)-\left\langle N(t_k)\right\rangle) \right\rangle \nonumber \\
    &\propto t_1^{\mu}\ldots t_k^{\mu} \;\;\frac{t_1}{t_k} \; .
    \label{eq:Covktimes}
\end{align}
To obtain these results, we rely on the assumption that for any values of the number of distinct sites visited $n_1$ and $n_2$ holds
\begin{align}
    \text{Cov}\left[\sum_{k=0}^{n_1-1}\tau_k,\sum_{k=n_1}^{n_2-1}\tau_k \right]=O\left( n_1^{2/\mu} \right) \; ,
     \label{eq:Criterion}
\end{align}
which is indeed verified for $1d$ L\'evy flights (see SI Sec. S5 B). In addition to the case of $1d$ L\'evy flights, where Eq.~\eqref{eq:Criterion} is satisfied, Eqs.~\eqref{cov} and \eqref{eq:Covktimes} provide in fact lower bounds on the correlation functions for recurrent RWs (see SI Sec.~S5.B for numerical checks),
\begin{align}
    &\left\langle  (N(t_1)-\left\langle N(t_1) \right\rangle )\ldots (N(t_k)-\left\langle N(t_k)\right\rangle) \right\rangle \nonumber \\
    &\geq  t_1^{\mu}\ldots t_k^{\mu} \;\;\frac{t_1}{t_k} \; .
    \label{eq:CovktimesIneq}
\end{align}
}
\blue{This lower bound is algebraically decreasing in $t_k$. The salient feature of these results is that temporal correlations in multiple-time distributions of recurrent RWs, such as those in Eq.~\eqref{cov}, have a long memory.}

\blue{Second, the distribution of $\tau_n$  allows us to provide a quantitative answer to the question raised in the introduction regarding  the disparity in life histories of foragers that starve if they do not eat after ${\cal S}$ steps. While in $1d$, the mean starvation time is known to increase linearly with  ${\cal S}$ (at large ${\cal S}$), the corresponding question in $2d$, which is relevant to most applications of foraging, is open. } \blue{We now show,} \blue{by relying on the results introduced in this paper, that the mean number of sites visited and consequently  the starvation time in $2d$ increases quadratically with ${\cal S}$ (up to logarithmic corrections). We start with the observation that, knowing that $n$ sites have been visited, the probability to starve is given by the probability that the time $\tau_n$ to visit a new site is larger than the metabolic time $\mathcal{S}$, $\mathbb{P}(\tau_n>\mathcal{S})=\sum_{\tau>\mathcal{S}}F_n(\tau)$. Using Table 1, we have that for $t_n=\sqrt{n}<\mathcal{S}$, the probability to starve is stretched exponentially small (up to algebraic prefactors), $\mathbb{P}(\tau_n>\mathcal{S})\approx \exp\big[-\sqrt{\mathcal{S}/t_n} \big]$. The desert (domain witout food) formed by the set of visited sites is too small to prevent the RW from finding new sites: the RW visits ${\cal S}^2$ sites in total in this first regime. However, for $t_n=\sqrt{n}>\mathcal{S}$, the probability for the RW to starve before finding a new site is large, as it is given by the tail of an algebraic distribution $\mathbb{P}(\tau_n>\mathcal{S})\propto 1/\mathcal{S}$. Consequently, the number of sites visited in this regime is negligible compared to the first one. Thus, the number of sites visited at starvation is given, up to log corrections,  by $n=\mathcal{S}^2$ and the lifetime by $\sum\limits_{k=1}^{\mathcal{S}^2} \left\langle \tau_k \right\rangle \sim \mathcal{S}^2$. This result is confirmed numerically in Figure~\ref{fig:Islands}{\bf f}. This resolves the open question of the lifetime of $2d$ starving random walks \cite{Benichou_Redner_2014,sanhedrai2021,sanhedrai2020lifetime,benichou2016role,krishnan2018optimizing}.
%It would be of particular interest to test this on real life organisms, for example by comparing lifetimes when they are located in a $1d$ object like a tube or a $2d$ one like a petri dish. 
}

\blue{Finally, the generality of our results opens the question of extending them to the challenging situation of non-Markovian processes, which is  a priori not covered by our approach. 
However, we argue in SI Sec S5.C that our results concerning the recurrent case can be extended to non-Markovian processes.
%However, we argue in SI that our results concerning the recurrent case can be extended to  non Markovian processes, up to the replacement $\mu\to 1-\theta$ (see SI S5 XXX) \footnote{\leo{For Markovian processes, $\theta=1-\mu$ (cite XXX).}}, where $\theta$ is the famous persistence exponent, defining the decay of the survival probability in presence of a target point \cite{Schehr_2013,Levernier2018}.  
 The agreement with numerical simulations of  highly  non-Markovian processes such as  the Fractional Brownian Motion \cite{hosking1984} (in the sub- and super-diffusive cases) and the True Self Avoiding Walk \cite{Amit} (see SI Sec S5.C.{\it4} for definition) is displayed in Figure~\ref{fig:Islands} ({\bf g}, {\bf h} and {\bf i} respectively). We point out again that this behavior $F_n(\tau)\propto \tau^{-(2-\theta)}$ is in sharp contrast to the usual decay of  the  first-passage probability to a target $F_{\rm target}(\tau)\propto \tau^{-(1+\theta)}$. This difference originates both from the complex geometry of the support of the random walk and potential memory, which, remarkably, are  universally  accounted for by our results.   } 

We have shown that the times between successive visits to new sites are a fundamental and useful characterization of the territory explored by a RW. We identified three temporal regimes for the behavior of these inter-visit time distributions, as well as the physical mechanisms that underlie these different regimes. In addition to their fundamental nature, these inter-visit times satisfy strikingly universal statistics, in spite of the geometrical complexity of the support of the underlying RW processes. The elucidation of these inter-visit times represents a promising research avenue to discover many more aspects of the intriguing exploration dynamics of RWs, as shown by the first applications provided here in the case of non-Markovian processes.

%\leo{even for strongly non-Markovian ones as supported by first results presented in SI.}

\section*{Methods}

\hspace{-0.5cm} {\it (i) Numerical simulations of recurrent and marginal RWs:} %The recurrent case is illustrated by: 
\begin{itemize}[leftmargin=0.3cm]
    \item L\'evy flights in $1d$ with $\alpha\in [1,2[$, where the jump length %and direction 
    is drawn from ${p(s)=1/[2\zeta(1+\alpha)|s|^{1+\alpha}]}$. Intermediary sites between initial and final positions of the jump are not visited.
    \item  Nearest-neighbour RWs on the Sierpinski gasket. The gasket is unbounded, and each neighbouring site is chosen with equal probability. Each RW starts at the central site (red square on Figure~\ref{fig:All}\textbf{b}).
    \item Nearest-neighbour RWs on the T tree. The T tree is generated up to generation 9, and then we perform a RW starting at the central site. Each neighbouring site is chosen with equal probability.
    \item Nearest-neighbour RWs on critical percolation clusters. The clusters are constructed from a ${1000\times1000}$ periodic square lattice, from which half of the bonds were randomly removed and then the largest cluster was selected. We start from a site chosen uniformly on the cluster. Each neighbouring site is chosen with equal probability.
    \item Nearest-neighbour RWs on the $2d$ lattice, persistent and not persistent. For persistent random walk, the probability to do the same step as the previous one is larger than $1/4$, while the probability to go in any other direction is taken uniformly among the 3 directions left.
\end{itemize}
We perform the random walks to get the domain $\bold{r}_n$ of $n$ distinct visited sites. To obtain the time $\tau_n$ to visit a new site, we use the exact enumeration method based on the adjacency matrix $M(\bold{r}_n)$ of the visited domain. The $\theta_n$, based on which the rescaled data lead to Figure~\ref{fig:All}, are obtained by measuring the slope of the exponential decrease at large times of the statistics of $\tau_n$.\\
{\it (ii) Numerical simulations of transient RWs:} In addition to the exact enumeration used to obtain the exit time statistics from the visited domain $\bold{r}_n$, we rely on a Monte Carlo Markov Chain generation of $\bold{r}_n$ on hypercubic lattices $d=3,$ $4$, $5$ and $6$ (we generate the visited domains in the same way as for recurrent RWs for the persistent RW in $3d$ or transient L\'evy flights). Using the observation that the average exit time is proportional to the surface of the visited domain, we bias the generation of these domains towards states of small surfaces. The bias is generated by a Wang-Landau procedure, in order to obtain a uniform probability on the surface of the visited domain, resulting in an increased probability of the small surface states.
\\
{\it (iii) Numerical simulations of non-Markovian RWs:} For the True Self Avoiding Random Walk on the $1d$ line, we record the number of visits $C_i$ of site $i$. The probability to jump to the site on the right is given by $\exp(-C_{i+1})/(\exp(-C_{i+1})+\exp(-C_{i-1}))$, otherwise the RW jumps on the left. For the fractional brownian motion (fBM), we use the module fbm \cite{fbmModule} of python based on Hosking’s method \cite{Hosking}. %This gives a list of times and positions of the random walker at these times. 
We discretize the line in intervals of size one, and consider that an interval has been visited when the RW enters it for the first time. $\tau_n$ is the time elapsed between visit of the $n^{\rm th}$ interval and the new $(n+1)^{\rm st}$ interval.

\section*{Data availability}
 The data generated in this study have been deposited in a GitHub repository \cite{codeUniv} located at: \url{https://github.com/LeoReg/UniversalExplorationDynamics.git}.

\section*{Code availability}
The code used to generate the data presented in this study have been deposited in a GitHub repository \cite{codeUniv} located at: \url{https://github.com/LeoReg/UniversalExplorationDynamics.git}.

%\newpage
%\clearpage

%\bibliography{refs}
%\bibliographystyle{apsrev4-1}


\section*{Acknowledgements}
S.R. gratefully acknowledges partial financial support from NSF grant DMR-1910736. We thank A.K. Hartmann, P. Viot and J. Klinger for useful discussions.

\section*{Author Contributions}
O.B., L.R. and M.D.   contributed to analytical calculations. L.R., M.D. and S.R. performed numerical simulations. All the authors  wrote the manuscript. O.B. conceived the research.

\section*{Competing Interests}
The authors declare no competing interests.

\end{document}

\RestyleAlgo{ruled}
\title{SUPPLEMENTARY INFORMATION\\
Universal exploration dynamics of random walks}

\author{L\'eo R\'egnier}
%\address{Laboratoire de Physique Th\'eorique de la Mati\`ere Condens\'ee, CNRS/Sorbonne University, 4 Place Jussieu, 75005 Paris, France}
\author{Maxim Dolgushev}
%\address{Laboratoire de Physique Th\'eorique de la Mati\`ere Condens\'ee, CNRS/Sorbonne University, 4 Place Jussieu, 75005 Paris, France}
\author{S. Redner}
%\address{Santa Fe Institute, 1399 Hyde Park Road, Santa Fe, NM, USA 87501}
\author{Olivier B\'enichou}
%\address{Laboratoire de Physique Th\'eorique de la Mati\`ere Condens\'ee,CNRS/Sorbonne University, 4 Place Jussieu, 75005 Paris, France}

\maketitle

\tableofcontents

\section{Scaling form of the distribution $F_n(\tau)$ for a nearest-neighbor random walk in $1d$}
\label{sec:1d}
In $1d$, $\tau_n$ is the exit time from an interval of length $L=n$ starting at a distance $a=1$ from the edge of the interval. The Laplace transform in the continuous limit is well known \cite{redner_2001}, and given by
\begin{eqnarray}
\widehat{F}_L(s) &=&\frac{\sinh\left(\sqrt{\frac{s}{D}}a\right)+\sinh\left(\sqrt{\frac{s}{D}}(L-a)\right)}{\sinh\left(\sqrt{\frac{s}{D}}L \right) }\;,
\end{eqnarray}
where $D$ is the diffusion coefficient. Then in the limit $L\to\infty$, $s\to0$ with $sL^2$ fixed, we obtain 
\begin{equation}
    \widehat{F}_L(s)\sim1-\sqrt{\frac{s}{D}}a\tanh\left(\sqrt{\frac{s}{D}}\frac{L}{2}\right) \; .
\end{equation}
By using residue calculus, and replacing $a$ by 1, $L$ by $n$, and setting $D=1/2$, we obtain
\begin{equation}
\label{function_1D}
        F_n(t)\sim\frac{2\pi^2}{n^3}\sum_{k=0}^\infty\left(2k+1\right)^2\,
        e^{-\pi^2\left(2k+1\right)^2t/2n^2}\;,
\end{equation}
which we may re-express in the following scaling form as written in the introduction of the main text,
\begin{equation}
\label{scaling1D}
    F_n(t)\sim n^{-3}\psi\left(\frac{t}{n^2}\right) \; .
\end{equation}

\section{Distribution $Q_n(r)$ of the  radius of the largest spherical region free of traps after $n$ sites have been visited}
\label{sec:rhon}

We determine the typical value $\rho_n$ of the radius of the largest spherical region completely covered by a RW, when $n$ distinct sites have been visited. We make use of the result for the mean of the number of distinct sites visited $N(t)$ at time $t$ by a RW (neglecting prefactors of order 1) \cite{Hughes_1995,gillis_2003}:
\begin{align}
\label{eq:scaling}
 \left\langle N(t)\right\rangle=n\sim
\begin{cases}
t^\mu,  & \qquad\mu\equiv {d_\text{f}}/{d_\text{w}}<1 \quad~\text{ (recurrent RWs)} \\
t/\ln t,       & \qquad\mu=1  \qquad\qquad\quad\text{ (marginal RWs)} \\
t,                   &\qquad  \mu>1  \qquad\qquad\quad\text{ (transient RWs)} \;.
\end{cases}
\end{align}

\subsection{Typical radius $\rho_n$ for marginal RWs }
\label{sec:cut-off}
\subsubsection{Theoretical argument}
We start with the marginal case, and follow the approach of  \cite{dembo2007large}. There, it was shown that the size of the largest ball entirely covered by a $2d$ random walker after $t$ steps typically behaves as $\rho_t=t^{1/4}$ (up to logarithmic prefactors). As we will show, this result can be extended to any marginal random walk and is given by $\rho_t=t^{1/2d_\text{f}}$ (up to logarithmic prefactors). This result is equivalent to the statement that the radius $r$ of the largest ball entirely covered by a marginal random walker before exiting this ball is typically $\sqrt{r}$, using that $r\sim t^{1/d_\text{w}}$ by the definition of $d_\text{w}$.

\begin{figure}[th!]
    \centering
    \includegraphics[width=0.25\columnwidth]{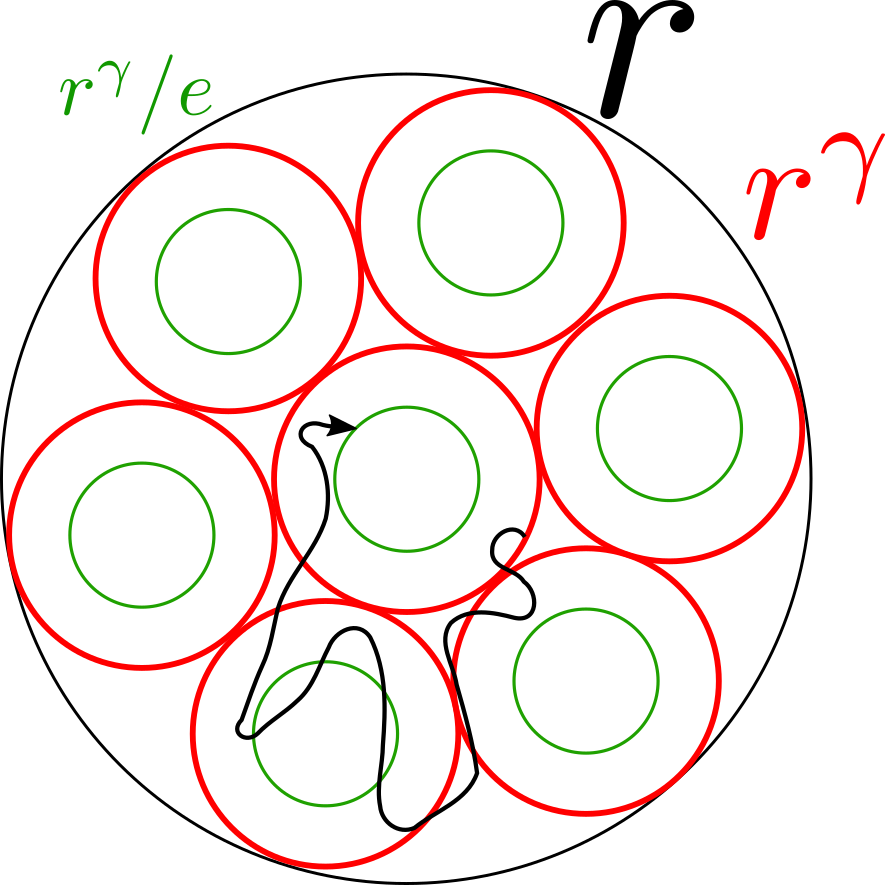}
    \caption{Representation of a $\gamma$-incursion in $2d$ (black curve with an arrow): a trajectory starting from a red sphere of radius $r^\gamma$ reaching the green sphere of radius $r^\gamma/e$ within the same sphere before the black sphere of radius $r$. }
    \label{fig:gammaincursion}
\end{figure}

The strategy is to split the ball of radius $r$ into disjoint balls of radius $r^\gamma$ with $0<\gamma<1$, and determine  the largest exponent $\gamma$ such that at least one ball of radius $r^\gamma/e^2$ \footnote{The prefactor $1/e^2$ is here only to ease comparison with numerical simulations.} has been fully visited before exiting the ball of radius $r$. Consider a ball of radius $r$ (\ref{fig:gammaincursion}). It contains $r^{d_\text{f}-\gamma d_\text{f}}$ disjoint balls of radius $r^\gamma$. First, we ask how many times the random walk will make an incursion inside a ball of radius $r^\gamma$ before exiting the ball of radius $r$. To answer this question, we use that the splitting probability of hitting a circle of radius $a$ before hitting the circle of radius $b$, when starting at distance $\rho$ from the center, is \cite{redner_2001}
\begin{align}
    \pi (\rho ) =\frac{\ln(\rho/b)}{\ln(a/b)}  \; . \label{eq:split2D}
\end{align} 
We say that the walker, which starts from the surface of the ball of radius $\rho=r^\gamma$, makes a $\gamma$-incursion inside the ball of radius $r^\gamma$ whenever it hits the ball of radius $a=r^\gamma/ e$ before a ball of radius $b=r$ (\ref{fig:gammaincursion}). The probability of such an incursion is \begin{align*}
  \frac{\ln(r^\gamma/r)}{\ln(r^\gamma/(e\,r))}\sim 1-\frac{1}{(1-\gamma)\ln r }\;.  
\end{align*} 
Thus the probability of making $k$ such incursions is 
\begin{equation}
    \mathbb{P}(X=k)= \left( 1-\frac{1}{(1-\gamma)\ln r}\right)^{k} \sim \exp\; \left( -\frac{k}{(1-\gamma)\ln r} \right)
\end{equation}
for one ball of radius $r^\gamma$.  That is, the distribution of the number of incursions is exponentially distributed.

Now we  assume  that the number of $\gamma$-incursions 
for all of the $M=r^{d_\text{f}-\gamma d_ \text{f}}$ balls are iid random variables $X_m$ ($m=1, \ldots, M$), that are exponentially distributed, with average $\left\langle X \right \rangle =(1-\gamma)\ln r$. Thus the average maximum number of $\gamma$-incursions $\mathcal{M}_M=\max X_m$ is given by
\begin{align}
    \left\langle \mathcal {M}_M \right\rangle &=\int_0^\infty \mathbb{P}\left(\mathcal {M}_M \geq k \right) {\rm d}k \nonumber \\
    &=\int_0^\infty \left(1- \mathbb{P}\left(X < k \right)^M \right) {\rm d}k \nonumber \\
    &= \int_0^\infty \left( 1-\left[1-\exp\left( -k\langle X\rangle \right) \right]^M \right) {\rm d}k \nonumber \\
     &= \int_0^\infty \sum_{i=0}^{M-1} \big[1- \exp\left(-k/\langle X\rangle\right)\big]^i\exp\left( -k/\langle X \rangle \right) {\rm d}k\nonumber \\
     &= \left\langle X \right \rangle \int_0^1 \sum_{i=0}^{M-1} \left(1- x \right)^i {\rm d}x 
%     &= \sum_{i=0}^{M-1}  \frac{\left\langle X \right \rangle}{i+1} \\
     \sim \left\langle X \right \rangle \ln \left( M \right) \; .
     \label{eq:Max_rv}
\end{align}
Replacing $M$ and $\left\langle X \right \rangle$ by their values above, we find that there are at most $d_\text{f}(1-\gamma)^2(\ln r)^2$ incursions inside a ball of size $r^\gamma$.

We now ask how many incursions inside a ball of radius $r^\gamma$ are necessary to visit all the sites within the ball of radius $r^\gamma/e^2$. We start by computing the probability to reach the origin when performing a $\gamma$-incursion, considering that the incursion stops when the RW hits the sphere of radius $r^\gamma$ (at which an other $\gamma$-incursion may occur). Using Eq.~\eqref{eq:split2D} for the splitting probability for an outer radius $b=r^\gamma$, inner radius $a=1$, and the radius of the starting position $\rho =r^\gamma /e$ (start of the $\gamma$-incursion), we obtain the probability to reach the origin during this incursion to be $1/(\gamma \ln r)$. Thus the probability of not reaching the origin during any of $k$ $\gamma$-incursions is 
\begin{align*}
 \left(1-\frac{1}{\gamma \ln r} \right)^k \sim \exp \left( -\frac{k}{\gamma \ln r}\right)\;.
\end{align*} 
Because there are $(r^\gamma /e^2)^{ d_\text{f}}$ sites inside a ball of radius $r^\gamma/e^2$, we estimate the probability of having at least one non-visited site after $k$ incursions to be 

\begin{align}
&1-\mathbb{P}(\text{all sites are visited in the ball of radius $r^\gamma/e^2$})\nonumber\\
    &\approx 1-\mathbb{P}(\text{$0$ is visited})^{(r^\gamma/e^2)^{ d_\text{f}}}\nonumber\\
    &=1-\left(1-\exp\left(-\frac{k}{\gamma \ln r} \right)\right)^{(r^\gamma/e^2)^{ d_\text{f}}}\nonumber\\
    &\approx (r^\gamma/e^2)^{ d_\text{f}} \times \exp \left( -\frac{k}{\gamma \ln r}\right)=\frac{1}{e^{2d_\text{f}}}\exp \left( \gamma \,d_\text{f}\, \ln r-\frac{k}{\gamma \ln r}\right),
\end{align} 

by assuming that all $(r^\gamma/e^2)^{ d_\text{f}}$ sites are equivalent to the origin, and exploration of all these specific sites are independent events. This probability goes to zero for $k>k_c(r)=\gamma^2 d_\text{f}\,(\ln r)^2$. Thus $\gamma^2 d_\text{f}\,(\ln r)^2$ incursions are necessary to visit all the sites within the ball of radius $r^\gamma$. 

\begin{figure}[th!]
    \centering
    \includegraphics[width=\columnwidth]{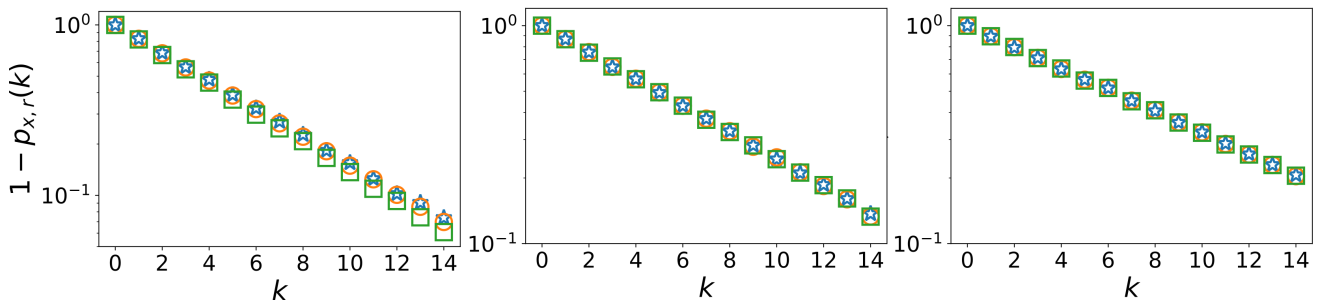}
    \caption{\blue{Distribution $1-p_x,r(k)$ of the probability not to have visited site $x$ among the $k$ $\gamma-$incursions ($\gamma=1$) inside a ball of radius $r=100/e$, $r=1000/e$ and $r=10000/e$ (from left to right), and for $x=0$, $x=10$, $x=20$ (blue stars, orange circles, and green squares, respectively).}}
    \label{fig:Equiv}
\end{figure}

Finally, since there are at most $(1-\gamma)^2\,d_\text{f} (\ln r)^2$ incursions inside a ball of radius $r^\gamma$ and that $\gamma^2\,d_\text{f}\,(\ln r)^2$ incursions are necessary to visit the entire ball of radius $r^\gamma/e^2$, equating these two numbers gives
\begin{equation}
    \gamma=1/2.
\end{equation}
Thus
\begin{equation}
    \rho_n=\sqrt{r}\sim t^{1/2d_\text{w}}\sim n^{1/2d_\text{f}} \; , \label{eq:mar_rho}
\end{equation}
up to logarithmic corrections.  This is the scaling behavior of $\rho_n$ for marginal RWs written in the main text.
\blue{
\subsubsection{Numerical check of the hypotheses.}
We check numerically the different assumptions which lead to the scaling obtained in the previous part, by relying marginal $1d$ L\'evy flights (for which $\rho_n\sim n^{1/2}$). We start by checking that the $\left(r^\gamma/e^2\right)^{d_\text{f}}$ sites are equivalent to the origin, in the sense that they have the same probability to be visited by the $k^\text{th}$ incursion as the origin. We compute numerically the probability that a site $x$ has been visited at the $k^\text{th}$ $\gamma-$incursion inside a domain of radius $r/e$ (taking $\gamma=1$, as here this exponent plays no role), starting at $\pm r/e$ and stopping the incursion when the ball hits the radius $r$. We denote this probability by $p_{x,r}(k)$.  Then, we check that $p_{x,r}(k)$ is independent of the site $x$. In \ref{fig:Equiv}, we take different values of $x$ and $r$ and verify that, indeed, at radius $r$ large enough, the visitation statistics at the $k^\text{th}$ incursion is independent of the site $x$.\\
}

\begin{figure}[th!]
    \centering
    \includegraphics[width=\columnwidth]{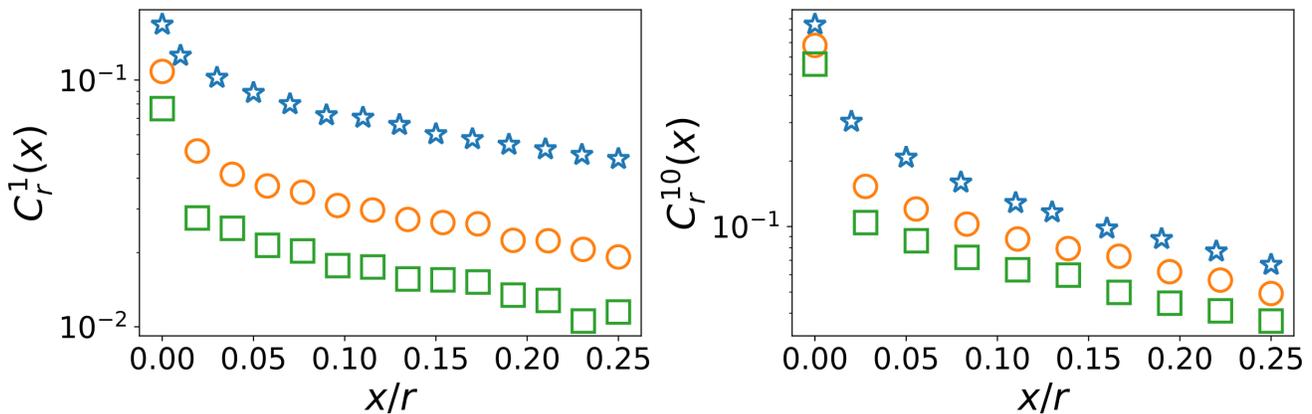}
    \caption{\blue{Relative difference $C_r^k(x)$ of the simultaneous visitation probability of site $0$ and site $x$ for $k=1$ (left) and $k=10$ (right), for $\gamma-$incursions ($\gamma=1$) inside a ball of radius $r=10^2/e$, $r=10^4/e$, and $r=10^6/e$ (blue stars, orange circles, and green squares, respectively). }}
    \label{fig:Indep}
\end{figure}

\blue{We then confirm the hypothesis on the independence of sites by measuring in numerical simulations the correlation in the visitation events. More precisely, we look at the probability of having visited site $0$ and site $x$ before  or at the $k^\textbf{th}$ $\gamma-$incursion inside a ball of radius $r/e$ ($\gamma=1$, the model is the same as for the first part of this answer and the exponent still plays no role). We denote this probability by $p_{0 \cap x, r}(k)$. In particular, we looked at the cases $k=1$ and $k=10$. Then, the independence is verified if $p_{0 \cap x, r}(k) \approx p_{0, r}(k)p_{x, r}(k)$. Because for large values of $k$ $p_{0 \cap x, r}(k) \approx p_{0, r}(k)\approx p_{x, r}(k) \approx 1$, we say that the visitation events are effectively independent when the relative difference
\begin{align}
    C^k_r(x)=\frac{|p_{0 \cap x, r}(k)-p_{0, r}(k)p_{x, r}(k)|}{1-p_{0 \cap x, r}(k)}
\end{align}
is smaller than a threshold value ($0.1$ for example). In \ref{fig:Indep}, we show that for increasing values of $r$, for both $k=1$ and $k=10$, the fraction of sites $x$ whose visitation is effectively independent of the visitation of site $0$ increases. This is a direct consequence of the decrease of $C^k_r(x)$ with both $x/r$ and $r$. \\
}

\blue{Finally, the equivalence and the independence hypotheses lead to the scaling $\rho_n \propto n^{1/2}$ for the marginal case.\\
Additionally, we check the scaling of $k_c(r)$ with $r$ in \ref{fig:kc}, $ k_c(r)$ being the mean number of incursions starting at $\pm r/e$ in a ball of radius $r$ necessary to have visited all the sites inside the ball of radius $r/e^2$. The $(\ln r)^2$ asymptotics is confirmed, even though the convergence occurs only at large values of $r$.
\begin{figure}[th!]
    \centering
    \includegraphics[width=0.5\columnwidth]{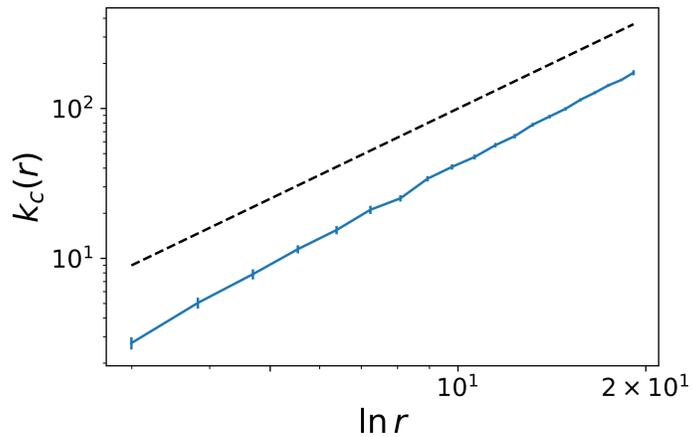}
    \caption{$k_c(r)$ compared to $(\ln r)^2$ in black dashed line for a $1d$ L\'evy flight of parameter $\alpha=1$. Errors bars are the $95\%$ confidence intervals.}
    \label{fig:kc}
\end{figure}
}
\subsection{Typical radius $\rho_n$ for recurrent RWs}
\label{sec:MaxRadTrans}

The splitting probability in the previous subsection is, in case of recurrent RWs~\cite{condamin2008,levernier2018universal},
\begin{align}
    1-\pi (\rho ) \propto \left(\frac{\rho}{b} \right)^{-\Delta}  \; , \label{eq:splitrecurrent}
\end{align} 
where, for notational simplicity, $\Delta\equiv d_\text{f}-d_\text{w}<0$
and $a$ is taken to 0 as for recurrent random walks the size of a target is not relevant to determine the probability of it being reached. Hence, the probability of making $k$ incursions in the ball of radius $r^\gamma$ is $\exp \left(-k r^{\Delta(1-\gamma)} \right)$. Thus the maximum number of incursions is $\ln(r^{d_\text{f}(1-\gamma)})\;r^{-\Delta(1-\gamma)}=d_\text{f}(1-\gamma)\; r^{-\Delta(1-\gamma)}\;\ln r$, by following similar steps to those used to derive Eq.~\eqref{eq:Max_rv}. 

Using Eq.~\eqref{eq:splitrecurrent} for the splitting probability for an outer radius $b=r^\gamma$ and the radius of the starting position $\rho =r^\gamma /e$, we obtain $\pi (\rho)\sim\text{const.}\equiv\pi_0$. The probability $\pi_0^k$ of not seeing the origin during any of the $k$ incursions is then independent of $r$.  Because there are $(r^\gamma/e^2)^{ d_\text{f}}$ sites inside the ball of radius $r^\gamma/e^2$, we estimate the probability of having at least one non-visited site when $k$ incursions have occurred as $(r^\gamma/e^2)^{ d_\text{f}} \times \pi_0^k $, by asserting that all sites are equivalent to the origin, and that the exploration of sites are independent events. This probability goes to 0 for $k>k_c(r)=\gamma d_\text{f}\, \ln r/\ln(1/\pi_0)$. Note that for large $r$, the number of incursions required before visiting every site ($\propto\gamma\ln r$) is much smaller than the expected maximum number of incursions in a ball ($\propto(1-\gamma)\; r^{-\Delta(1-\gamma)}\;\ln r$). This means that $\gamma=1$.  These considerations lead to  
\begin{align}
    \rho_n=r\sim t^{1/d_\text{w}}=n^{1/d_\text{f}}\; . \label{eq:rec_rho}
\end{align}

\subsection{Typical radius $\rho_n$ for transient RWs}
For transient walks, the splitting probability is~\cite{condamin2008,levernier2018universal}:
\begin{align}
    \pi (\rho ) =  1-\frac{{\displaystyle 1-\left(\frac{a}{\rho}\right)^{\Delta}}}
    {\displaystyle {1-\left(\frac{a}{b}\right)^{\Delta} } } \; . \label{eq:splitTrans}
\end{align}
When $\rho\rightarrow\infty$,
%Taking the values $a=r^\gamma/e$, $b=r$ and $\rho=r^\gamma$, the probability of making a $\gamma$-incursion becomes constant for large $r$, 
$\pi (\rho)\sim\text{const.}\equiv\pi_1$. Thus the probability of making $k$ incursions into a ball of radius $r^\gamma$ is $\pi_1^k$, and thus the maximum number of incursions is $\ln(r^{d_\text{f}(1-\gamma)})/\ln( 1/\pi_1)$. 

Using Eq.~\eqref{eq:splitTrans} for the splitting probability for an outer radius $b=r^\gamma$, inner radius $a=1$, and the radius of the starting position $\rho =r^\gamma /e$, we obtain the probability of not reaching the origin during any of these $k$ incursions: 
\begin{align*}
\left(1-\pi(r^\gamma /e)\right)^k=\left(1-\frac{e^{\Delta}-1}{r^{\gamma\Delta}-1}\right)^k \sim \exp \left(-k\frac{e^{\Delta}-1}{r^{\gamma\Delta}} \right)\;.
\end{align*}
Because we have to visit $(r^\gamma/e^2)^{ d_\text{f}}$ sites, the random walk must take 
\begin{align*}
k>k_c(r)=\gamma d_\text{f}\;\frac{r^{\gamma\Delta}}{e^{\Delta}-1}\, \ln r
\end{align*}
steps to ensure that all sites have been visited. We observe that the number of incursions to visit every site inside the ball of radius $r^\gamma/e^2$ is much larger than the expected maximum number of incursions into a ball ($\propto(1-\gamma)\ln r$ vs. $\propto\gamma r^{\gamma\Delta}\ln r$, respectively). This means that $\gamma=0$, and the typical size of a region entirely visited by a transient walk when exiting the domain of size $r$ is smaller than any power law in $r$ or in $n$, 
\begin{align}
    \rho_n=O(r^0)=O(n^0)\;.
\end{align}
Indeed, as we will show in the following, the radius of the largest ball for which every site is visited grows as $\rho_n \sim \left( \ln t \right)^{1/\Delta} \sim \left( \ln n \right)^{1/\Delta}$ (because the number of visited sites grows linearly with the number of steps, we do not distinguish between $\rho_n$ defined with $n$ the number of distinct sites visited or $\rho_t$ with $t$ the number of steps). First, we cut the trajectory of the RW of length $n$ into non-overlapping balls of radius $r$. There are approximately $n/r^{d_\text{w}}$ such balls, because the fractal dimension of the visited domain is $d_\text{w}$. For each of these balls, indexed by $i$, we are interested in the time $t_i$ that the RW spends inside the ball before exiting. Because the walk is transient, we suppose that once the RW exits the ball, it never returns. For each of the balls, the exit time is exponentially distributed (at long times) and its average is proportional to $r^{d_\text{w}}$ \cite{Hughes_1995}. Now we make the approximation that the exit times $t_i$ are independent, and we obtain that the maximal value for $t_i$ is given by $t_{\text{max}}(n)\propto r^{d_\text{w}}\ln \left( n/r^{d_\text{w}} \right)$, as a consequence of Eq.~\eqref{eq:Max_rv}. Using that the RW is transient, the number of distinct sites visited by the RW up to time $t_{\text{max}}(n)$ is proportional to $t_{\text{max}}(n)$. Thus at least one ball of radius $r$ has been fully visited as soon as  
\begin{align}
    t_{\text{max}}(n) > r^{d_\text{f}} \; .
\end{align}
Thus the optimal radius $\rho_n$, which corresponds to $r$ such that $t_{\text{max}}(n) = r^{d_\text{f}}$, is
\begin{align}
\rho_n = \left( \ln n \right)^{1/\Delta} \; . \label{eq:tra_rho}
\end{align}
Ref.~\cite{erdos1991} has proven this result for RWs with nearest-neighbor jumps ($d_\text{w}=2$) on a hypercubic lattices in dimension $d \geq 3$. Equation~\eqref{eq:tra_rho} is also compatible with the previous prediction that $\rho_n$ grows more slowly than any power of $n$. 

\subsection{Scaling form of $Q_n(r)$}
\begin{figure}[th!]
    \centering
    \includegraphics[width=\columnwidth]{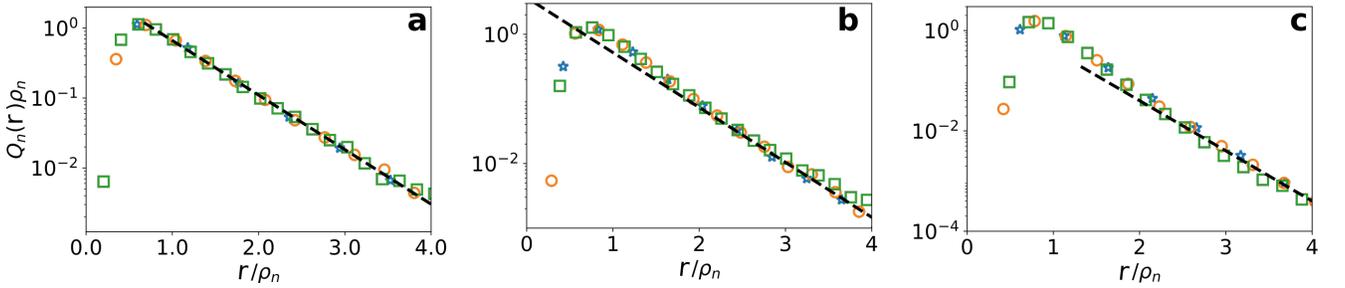}
    \caption{Distribution of the radius of the largest trap-free domain for L\'evy flights in $1d$, with the radius rescaled by the average value $\rho_n$ for: {\bf a} $\alpha=1.2$ ($n=1600,\; 3200, \; 6400$ in blue, orange, and green, respectively), {\bf b} $\alpha=1$  ($n=400,\; 800, \; 1600$ in blue, orange, and green, respectively), and {\bf c} $\alpha=0.8$ ($n=200,\; 800, \; 3200$ in blue, orange, and green, respectively).}
    \label{fig:recurrent_distrib_free}
\end{figure}
\begin{figure}[th!]
    \centering
    \includegraphics[width=\columnwidth]{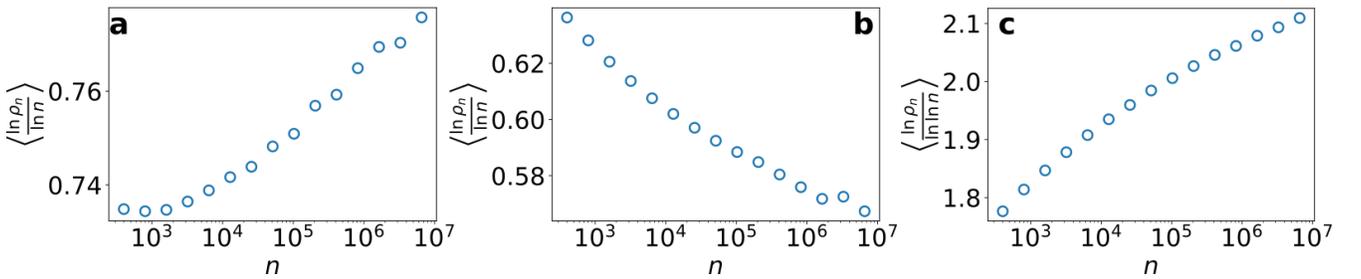}
    \caption{Evolution of $\rho_n$ with $n$, defined as the radius of the largest fully visited ball within a visited domain of $n$ sites, for L\'evy flights with parameters: {\bf a} $\alpha=1.2$, {\bf b} $\alpha=1$, and {\bf c} $\alpha=0.8$.}
    \label{fig:rho_n}
\end{figure}

From the typical length scale $\rho_n$ determined above, we argue that the probability density $Q_n(r)$ of the radius of the largest spherical trap-free region free is
\begin{align}
Q_n(r)=\frac{1}{\rho_n}f\left( r/\rho_n\right).
\end{align}
The characteristic length $\rho_n$ can be interpreted as a correlation length, so that two sites separated by a distance larger than $\rho_n$ have effectively independent visitation probabilities. Similar to the celebrated idea of independent ``blobs'' defined in~\cite{degennes,grosberg}, we consider that the RW represents a polymer that can be separated into independent blobs.  Within each blob of radius $\rho_n$, all the sites are either all visited or not visited at all, and the visitation probabilities between distinct blobs are independent. For large values of $r/\rho_n$, this leads to an exponential distribution:
\begin{align}
Q_n(r)\approx \frac{1}{\rho_n}\;e^{-a \left(r/\rho_n \right)^{d_\text{f}}} \; ,
\end{align} 
since $\left(r/\rho_n \right)^{d_\text{f}}$ is the number of independent domains free of traps/fully visited blobs. Here $a$ is a constant that is independent of $r$ and $n$. 

We confirm the scaling form as well as the asymptotic exponential decay in \ref{fig:recurrent_distrib_free}. \ref{fig:rho_n} displays the extraordinarily slow approach of $\rho_n$ to the asymptotic formulas derived above.  These results are illustrated for Lévy flights for the three main cases. For recurrent ($\alpha=1.2$), marginal ($\alpha=1$), and transient ($\alpha=0.8$) flights, we use Eqs.~\eqref{eq:rec_rho},~\eqref{eq:mar_rho},~\eqref{eq:tra_rho} to predict that $\ln \rho_n/\ln n\to 1$, $\ln \rho_n/\ln n\to 1/2$, and  $\ln \rho_n/\ln  \ln n\to 5$, respectively.

\section{Analytical expressions for $F_n(\tau)$ in Table I of the main text} \label{sec:FractalDim} 

\subsection{Short-time behavior of $F_n(\tau)$: algebraic decay}
\subsubsection{Equation (1) of the main text}
Similar to the approach given in \cite{Plyukhin2016}, we start by deriving a renewal equation for the exit-time distribution, conditioned on the visited domain consists of $n$ sites, whose list of positions is denoted by the vector $\mathbf{r}_n$.  We set the position of the RW  at $\tau=0$ to be $x$. For the list of previously visited sites $\mathbf{r}_n$ and starting position $x$, we denote $F_n(\tau', x' |\mathbf{r}_n,x)$ as the first-passage probability to site $x'$ at time $\tau'$ before any reaching other of the non-visited sites. The quantity $P_\text{trap}(\tau|\mathbf{r}_n,\; x)$ is defined as the probability to be at a non-visited site at time $\tau$, conditioned on the visited domain being $\mathbf{r}_n$ and the starting position of the RW is $x$ at time $\tau=0$. Partitioning this last event over the first-passage at site $x'$ at time $\tau'$, we obtain
\begin{align}
    P_\text{trap}(\tau|\mathbf{r}_n,\; x)=\delta (\tau) +\sum_{\tau'=1}^\tau \sum_{x'\notin \mathbf{r}_n} F_n(\tau', x' |\mathbf{r}_n,x)P_\text{trap}(\tau-\tau'|\mathbf{r}_n, x') \; .
\end{align}
Here we make the approximation that since $x$ is close to the surface of $\mathbf{r}_n$, we take the starting position $x$ to be at a trap. To perform averages over the initial site $x$, we introduce $P(\mathbf{r}_n,x)$ as the joint probability for the RW to visit the set of sites $\mathbf{r}_n$ and is at position $x$ at time $\tau=0$, so that the quantity $P_\text{trap}(\tau)$ we are interested in can be written as 
\begin{align}
    P_\text{trap}(\tau)&\equiv\left\langle P_\text{trap}(\tau|\mathbf{r}_n,\; x)  \right\rangle=\sum_{\mathbf{r}_n,x}P_\text{trap}(\tau|\mathbf{r}_n,\; x)P(\mathbf{r}_n,x)\;.
\end{align}
Using this definition, we have
\begin{align}
    P_\text{trap}(\tau)&=\delta (\tau) +\sum_{\tau'=1}^\tau \left\langle\sum_{x'\notin \mathbf{r}_n}  F_n(\tau', x' |\mathbf{r}_n,x)P_\text{trap}(\tau-\tau'|\mathbf{r}_n, x') \right\rangle \nonumber\\
    &\approx \delta (\tau) +\sum_{\tau'=1}^\tau \left\langle  F_n(\tau'|\mathbf{r}_n,x)P_\text{trap}(\tau-\tau'|\mathbf{r}_n) \right\rangle \nonumber \\
    &\approx \delta (\tau) +\sum_{\tau'=1}^\tau   F_n(\tau')P_\text{trap}(\tau-\tau')\;. \label{eq:MF}
\end{align}
In the mean-field type formula above, we neglected the dependence of $P_\text{trap}(\tau-\tau'|\mathbf{r}_n, x')\approx P_\text{trap}(\tau-\tau'|\mathbf{r}_n)$ on the starting site $x'$ (keeping in mind only that $x' \notin \mathbf{r}_n$) and the  spatial correlations in the visited domain. Numerical simulations (see early-time algebraic regime of Fig. 3\textbf{a}, \textbf{b}, and \textbf{c} of the main text and of \ref{fig:fractals}\textbf{a}) justify  the applicability of \eqref{eq:MF}. For early times $\tau$ and for visited domains of large size $n$, $F_n\approx F_\infty$, and we finally get Eq.~(1) of the main text.

\subsubsection{Solution to Eq.~(1) of the main text}

\begin{figure}
    \centering
    \includegraphics[width=\columnwidth]{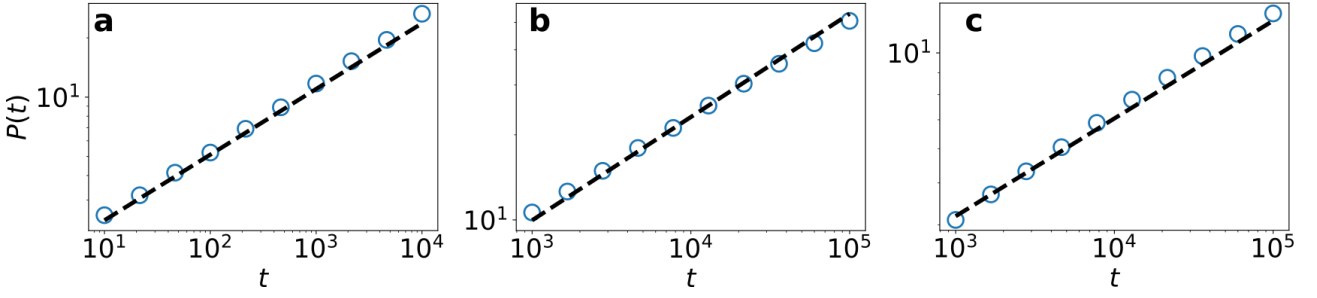}
    \caption{Average perimeter $P(t)$ of the visited domain for {\bf a} $1d$ L\'evy flights of parameter $\alpha=1.5$, {\bf b} nearest-neighbor random walks on a Sierpinski gasket {\bf c} nearest-neighbor random walks on a T-tree. The black dashed line represents the prediction \eqref{eq:Perimeter}.}
    \label{fig:Perimeter}
\end{figure}

To solve this equation, we first need to determine the time dependence of $P_\text{trap}(t)$. From Eq.~(2) of the main text, $P_\text{trap}(t) \propto t^{(d_\text{T}-d_\text{f})/d_\text{w}}$, where $d_\text{T}$ is the  fractal dimension of the interface between visited and non-visited sites. The problem is then reduced to specifying $d_\text{T}$. For this, we determine the time evolution of the number of sites $P(t)$ on the interface between visited and non-visited sites for recurrent and marginal RWs at time $t$. By interpreting the set of $\left\langle\tau_n\right\rangle$ as the average discovery rate of new sites, we relate the perimeter $P(t)$ of the visited domain (the number of traps) and the size of the visited domain $V(t)$:
\begin{align}
    \label{eq:rate}
    \frac{{\rm d} V(t)}{{\rm d}t}&\sim \frac{1}{\left\langle\tau_{V(t)}\right\rangle} \;.
\end{align}
Using Kac's lemma \cite{kac1947notion}, the average return time to the boundary of the visited domain is
\begin{align} 
\label{eq:Kac}
\left\langle\tau_{V(t)}\right\rangle \propto \frac{V(t)}{P(t)} \; .
\end{align}
Combining Eqs.~\eqref{eq:rate} and \eqref{eq:Kac} leads to 
\begin{equation}
    P(t)\sim \frac{{\rm d} }{{\rm d}t}\left( V(t)^2 \right) \; .
\end{equation}
Using the results of Eq.~\eqref{eq:scaling}, with $n=V(t)$, we get 
\begin{equation}
    P(t) \sim 
\begin{cases}    
        t^{2d_\text{f}/d_\text{w}-1} & \qquad  d_\text{f}<d_\text{w}\\
        t/ \ln^2 t & \qquad d_\text{f}=d_\text{w}\;,
\end{cases}
\label{eq:Perimeter}
\end{equation}
which gives the fractal dimension of the surface (neglecting logarithmic prefactors):
\begin{align}
    d_\text{T}=2d_\text{f}-d_\text{w}, & \quad \mbox{ for } d_\text{f} \leq d_\text{w} \; .
\end{align}
We verify the time dependence of the perimeter of the visited domain in \ref{fig:Perimeter} for three types of recurrent random walks. We note that the result for the $2d$ random walk is already known from \cite{van1997universal} and for $1d$ L\'evy flights as well \cite{Mariz2001}. We note that in Kac's lemma the perimeter of the domain $D(t)$ of volume $V(t)$ would be defined as 
\begin{align}
    P(t)=\sum_{i \notin D(t)} p(i \to j) \; .
\end{align}
For nearest neighbour random walks it is straightforwardly proportional to the perimeter defined in the usual manner. For L\'evy flights, we verify that indeed it is similar by using this definition for the perimeter in \ref{fig:Perimeter}{\bf a}, and see that the scaling is the same as in \cite{Mariz2001}.
Using this result, we obtain the scaling behaviour  $P_\text{trap}(t)\propto t^{\mu-1}$.

We now relate $P_\text{trap}$ and $F_\infty$, which is most conveniently done in the Laplace domain. Defining the Laplace transform of a function $f(t)$ by $\mathcal{L}\left\{f(\tau)\right\}\equiv\widehat{f}(s)=\int_0^\infty f(t)e^{-st}dt$, Eq.~(1) of the main text leads to
\begin{align*}
    \widehat{P}_\text{trap} (s)=1+\widehat{F}_\infty(s)\widehat{P}_\text{trap} (s),
\end{align*}
which yields
\begin{align}
    \label{eq:Renew}
    \widehat{F}_\infty(s)=1-\frac{1}{\widehat{P}_\text{trap} (s)} \; .
\end{align}
Using $P_\text{trap}(t)\propto t^{\mu-1}$, we have $\widehat{P}_\text{trap} (s) \propto 1/s^\mu$. 

Finally, differentiating Eq.~\eqref{eq:Renew} with respect to $s$ (to get a diverging quantity as $s\to 0$), we obtain 
\begin{align*}
  \mathcal{L}\left\{\tau F_\infty(\tau)\right\}
    =-\frac{{\rm d}\widehat{F}_\infty(s)}{{\rm d}s}=\frac{{\rm d}}{{\rm d}s}\left(\frac{1}{\widehat{P}_\text{trap} (s)} \right)\propto s^{\mu-1}\; .
\end{align*}
From the Tauberian theorem~\cite{klafter2011first} that links the small-$s$ asymptotics in the Laplace domain to the long-time asymptotics, we finally obtain the algebraic decay of $F_\infty(\tau)$ that is written in the main text, 
\begin{align}
    F_\infty(\tau)\propto \tau^{-1-\mu} \; .\label{eq:Falg}
\end{align}

\subsubsection{The exponent of the algebraic decay: The first moment approach}  \label{sec:Direct}

We present here an alternative approach that allows one to recover the exponent $\delta$ that characterizes  the algebraic decay of $F_n\propto \tau^{-1-\delta}$ determined in Eq.~\eqref{eq:Falg}. This approach consists of identifying two expressions for $\left\langle \tau_n \right\rangle$.  On the one hand, the very definition of  $\left\langle \tau_n \right\rangle$ gives
\begin{align}
\left\langle \tau_n \right\rangle &\propto \int_1^{t_n} \frac{\tau }{\tau^{1+\delta}}\;{\rm d}\tau 
\propto t_n^{1-\delta} = n^{(1-\delta)/\mu} \; \label{eq:tau_n1}.
\end{align}
On the other hand, knowing that $\left\langle \tau_n \right\rangle$ is the time scale needed to discover a new site, 
\begin{align}
\label{eq:tau_n2}
\frac{{\rm d}\left\langle N(t) \right\rangle}{{\rm d}t}\sim t^{\mu-1}\sim \frac{1}{\left\langle \tau_{n=\left\langle N(t) \right\rangle} \right\rangle}.
\end{align}
Using finally the temporal scaling of the number of distinct sites visited (see Eq.~\eqref{eq:scaling}), $n\sim t ^{\mu}$, we again find $\delta=\mu$.

\subsubsection{Prefactor of the algebraic decay}
\label{subsec:Amu}
This subsection is devoted to the analysis of the prefactor $A(\mu)$ in the algebraic decay that appears in Fig.~3 of the main text.

\textit{Exact value for marginal RWs}. In the  marginal case ($\mu=1$), we can obtain the prefactor exactly.  We exploit the fact that there are two independent ways to express $\langle \tau_n\rangle$. The first method relies on the 
self-averaging property of $N(t)$ in which $\text{Var}(N(t))/\langle N(t)\rangle ^2 \to 0$ as $t\to\infty$ (see \cite{Mariz2001,Hughes_1995}). Thus the number of distinct sites visited any marginal RW is close to its average value, which scales asymptotically as $\beta t/\ln t$, with $\beta$ depending on the specific RW model. We now use the link between the distributions of $N(t)$ and $\sum\limits_{k \leq n}  \tau_k$:
\begin{align}
    \mathbb{P}\left(N(t) \geq n \right)=\mathbb{P}\Big( \sum\limits_{k \leq n}  \tau_k \leq t\Big)&\approx \Theta \left(\beta t/\ln t-n\right) \approx \Theta \left(t-n \ln n/\beta \right) \; ,
\end{align}
with $\Theta$ the Heaviside function, which is 0 for negative argument and 1 otherwise. This last equation shows that  $\sum\limits_{k \leq n}  \tau_k$  also has negligible fluctuations and is peaked at the value $(n \ln n)/\beta=\sum\limits_{k \leq n} \left\langle  \tau_k \right\rangle$. Thus
\begin{equation}
    \left\langle\tau_n\right\rangle \sim \frac{\ln n}{\beta}\;.
\end{equation} 

The second method relies on the algebraic decay $F_n(k)=A(\mu=1)k^{-2}$ for $k<t_n$, to 
compute the asymptotic behavior of the first moment:
\begin{align}
    \left\langle \tau_n \right\rangle =\sum_{k=1}^{\infty} kF_n(k)
    \sim \sum_{k=1}^{t_n} 
    \frac{A(1)}{k}+O(1) 
    \sim A(1) \ln t_n 
    \sim \tfrac{1}{2} A(1) \ln n\;,
\end{align}
where we used that $t_n \propto \sqrt{n}$ at large times. We note that the times $k >t_n$ do not contribute to the average, because of the rapid stretched exponential decay of $F_n(k)$. Identifying the expressions for $ \left\langle \tau_n \right\rangle$ in the previous two equations, we obtain the exact value 
\begin{equation}\label{eq:A1}
    A(1)=\frac{2}{\beta}
\end{equation} 
in the marginal case. 

We now use the known facts that $\beta=\pi$ for the nearest-neighbor random walk on the square lattice~\cite{gillis_2003} and  $\beta=3$ for L\'evy flights in $1d$ and parameter $\alpha=1$ \cite{gillis_2003,Mariz2001}. Thus we obtain $F_n(\tau)$, including the correct amplitude, for marginal RWs for $1 \ll \tau_n \ll \sqrt{n}$:
\begin{equation}
\label{resultsurface-F}
   F_n(\tau) \sim \left\{
    \begin{array}{ll}
        {\displaystyle \frac{2}{\pi\tau^2}} &\quad \mbox{for nearest-neighbor RWs on square lattice}\\[2mm]
         {\displaystyle  \frac{2}{3\tau^2}} &\quad  \mbox{for L\'evy flights in $1d$, $\alpha=1$\;.}
    \end{array}
\right.
\end{equation}

\textit{Recurrent RWs.} We obtain an approximate value of the prefactor $A(\mu)$ in the algebraic regime from the normalization condition
\begin{align}
    1&=\sum\limits_{k=1}^{\infty} F_n(k) =\sum\limits_{k=1}^{t_n}\frac{A(\mu)}{k^{1+\mu}}+\sum\limits_{k>t_n}\frac{\text{const}}{n^{1/\mu+1}}\;e^{-\text{const~}k/n^{1/\mu}} \sim A(\mu)\zeta(1+\mu)+O\left( \frac{1}{n}\right) \; \label{eq:Aeval},
\end{align}
from which we infer $A(\mu)=1/\zeta(1+\mu)$.

\subsection{Long-time behavior of $F_n(\tau)$: exponential and stretched exponential regimes}
\label{sec:stretched}
Here we obtain the stretched exponential and exponential decays at long times. We start from Eq.~(3) of the main text, written for $\tau>t_n$, 
\begin{eqnarray}
\label{bound}
S_n(\tau)\ge \frac{q_n}{\rho_n} \int_0^{n^{1/d_\text{f}}}  e^{-b \tau/r^{d_\text{w}}-a (r/\rho_n)^{d_\text{f}}} {\rm d}r\; ,\label{eq:eq3main}
\end{eqnarray}
where $q_n\equiv \int_{t_n}^\infty F_\infty(\tau') {\rm d}\tau'$ and $a$ and $b$ are constants. To determine the large-$\tau$ behavior of the integral on the rhs of Eq.~\eqref{eq:eq3main}, we apply the Laplace method. It is convenient to first make the variable change variable $r=\rho\, \tau^{\mathcal{D}}$, where again, for notational simplicity, $\mathcal{D}\equiv 1/(d_\text{w}+d_\text{f})$, to give
\begin{align}
S_n(\tau)&\ge \frac{q_n\tau^{\mathcal{D}}}{\rho_n} \int_0^{n^{1/d_\text{f}}/\tau^{\mathcal{D}}}  \exp\left[-\tau^{d_\text{f}\,\mathcal{D}}\;U(\rho)\right] {\rm d}\rho\; ,
\end{align}
with $U(\rho)\equiv b/\rho^{d_\text{w}}+a (\rho/\rho_n)^{d_\text{f}}$. To proceed with the Laplace method, the location of the minimum of $U(\rho)$, $\rho^*\sim\rho_n^{d_\text{f}\,\mathcal{D}}$ must be smaller than $n^{1/d_\text{f}}/\tau^{\mathcal{D}}$. Equivalently, $r^*(\tau)=\rho^*\,\tau^{\mathcal{D}}\sim \rho_n(\tau/t_n)^{\mathcal{D}}<n^{1/d_\text{f}}$.  This condition means that $\tau< n^{1+1/\mu}/\rho_n^{d_\text{f}}\equiv T_n$. Thus, for times $t_n<\tau<T_n$, the Laplace method gives
\begin{align}
S_n(\tau)&\ge \frac{q_n\tau^{\mathcal{D}}}{\rho_n} \sqrt{\frac{2\pi} {U''(\rho^*)\tau^{\mu/(1+\mu)}} }\;\exp\left[-\tau^{\mu/(1+\mu)}\;U(\rho^*)\right]\; .
\end{align}
Because 
\begin{align*} \rho^* \sim \rho_n^{d_\text{f}\,\mathcal{D}}=\rho_n^{\mu/(1+\mu)} \qquad\text{and} \qquad U(\rho^*)\sim \rho_n^{-\mu d_\text{w}/(1+\mu)}=t_n^{-\mu/(1+\mu)}\;,
\end{align*} 
we obtain for $t_n<\tau<T_n$:
\begin{align}
S_n(\tau)&\ge \text{const }\times q_n\tau^{\mathcal{D}-\mu/2(1+\mu)}\rho_n^{(d_\text{f}-2)/2(1+\mu)}\; \exp\left[-\text{const}\left(\tau/t_n\right)^{\mu/(1+\mu)}\right] \; .
\end{align}
For $\tau>T_n$ and $\rho$ in the range $[0, n^{1/d_\text{f}}/\tau^{1/(d_\text{w}+d_\text{f}})]$, $U(\rho)$ has its minimum at the upper bound $n^{1/d_\text{f}}/\tau^{\mathcal{D}}$. Thus we are left with 
\begin{align}
S_n(\tau)&\ge q_n \; \exp\left[-b\tau/n^{1/\mu}-an/\rho_n^{d_\text{f}}\right] \; .
\end{align}

Now we study the explicit form of this lower bound for each class of RWs  (recurrent, marginal and transient), following the results of section \ref{sec:rhon} about the scalings of $\rho_n$ with $n$. In addition, as in the classic trapping
problem, we expect that this lower bound for the survival probability will have the same time dependence as
the survival probability itself.
\begin{itemize}
    \item recurrent RWs.  Here $\mu<1$ and $\rho_n\sim n^{1/d_\text{f}}$. Because $t_n=n^{1/\mu}$ and $T_n=n^{1+1/\mu}/\rho_n^{d_\text{f}}=n^{1/\mu}$, there is no intermediate regime and we have
    \begin{align}
    S_n(\tau)&\approx q_n \;\exp\left[-b\tau/n^{1/\mu}-a\right] \; ,
    \end{align}
with $q_n=\int_0^{n^{1/\mu}}{\rm d}\tau/\tau^{1+\mu} \propto 1/n$. This leads, for $\tau\gg t_n$, to
    \begin{align}
    F_n(\tau)&\propto \frac{1}{n^{1+1/\mu}}  \exp\left[-b\tau/n^{1/\mu}\right] \; .\label{eq:Fexp}
    \end{align}
    \item marginal RWs. Here $\mu=1$ and $\rho_n\sim n^{1/2d_\text{f}}$, so that $t_n=\sqrt{n}$,  $T_n=n^{3/2}$ and $q_n\propto 1/\sqrt{n}$. Thus we expect for the intermediate regime (up to algebraic prefactors in $\tau$ and $n$) 
    \begin{align}
S_n(\tau)&\approx  e^{-\text{const}\times \sqrt{\tau/\sqrt{n}}} \; ,
\end{align}
which results in the stretched exponential behavior of Table I. For the exponential regime, we get 
\begin{align}
F_n(\tau)&\propto \frac{1}{n^{3/2}} \; e^{-b\tau/n-a\sqrt{n}} \; .
\end{align}
\item transient RWs. Here $\mu>1$ and $\rho_n\sim 1$ (neglecting logarithmic corrections).  This behavior for $\rho_n$ leads to $t_n=1$ (i.e., there is no algebraic regime), $T_n=n^{1+1/\mu}$, and $q_n\propto 1$. We expect for the intermediate regime (up to algebraic prefactors in $\tau$ and $n$) 
    \begin{align}
S_n(\tau)&\approx  e^{-\text{const}\times\tau^{\mu/(1+\mu)}} \; ,
\end{align}
which results in the stretched exponential behavior of Table I. For the exponential regime, we get 
\begin{align}
F_n(\tau)&\propto \frac{1}{n^{1/\mu}} \; \exp\left[-b\tau/n^{1/\mu}-an\right] \; .
\end{align} 
\end{itemize}

\subsection{Scaling form $F_n(\tau)$ for recurrent RWs, Eq.~(4) of the main text}

Here we show that the distribution $F_n(\tau)$ for recurrent RWs admits the scaling form  Eq.~(5) of the main text. As shown above, $F_n(\tau)$ displays two regimes for recurrent RWs: algebraic (Eq.~\eqref{eq:Falg}) and exponential (Eq.~\eqref{eq:Fexp}). Noting that the algebraic regime can be rewritten as $F_n(\tau)\propto \tau^{-1-\mu}=\frac{1}{n^{1+1/\mu}}\left(\tau/n^{1/\mu}\right)^{-1-\mu}$, both regimes can be synthesized into the  scaling form:
    \begin{equation}
    \label{scalingrecurrent}
    F_n(\tau)=\frac{1}{n^{1+1/\mu}} \;\psi \left( \frac{\tau}{n^{1/\mu}}\right )  \; .
    \end{equation}
The scaling function $\psi(x)$ decays algebraically for small $x$ and exponentially for large $x$.

\subsection{Moments of $\tau_n$} \label{sec:moments}
\subsubsection{Analytical derivation }
Moments of $\tau_n$ follow from Table I of the main text.
For recurrent walks ($\mu<1$), the scaling form \eqref{scalingrecurrent} yields
\begin{align}
    \left\langle\tau_n^k \right\rangle&=\sum_{\tau=1}^\infty \frac{\tau^k}{n^{1/\mu+1}}\psi\left( \frac{\tau}{n^{1/\mu}} \right) \sim n^{k/\mu-1}\int_0^\infty u^k \psi(u) {\rm d}u \; .
\end{align}

For marginal walks ($\mu=1$), we use the result that the algebraic regime holds up to $\tau \propto t_n=\sqrt{n}$ and gives the scaling behavior of the moments as
\begin{align}
    \left\langle\tau_n^k \right\rangle&\propto \sum_{\tau=1}^{\sqrt{n}} \frac{\tau^k}{ \tau^2}= \sum_{\tau=1}^{\sqrt{n}} \tau^{k-2} \; ,
\end{align}
which results in $\left\langle\tau_n \right\rangle\propto\ln n$ and $\left\langle\tau_n^k \right\rangle\propto n^{(k-1)/2}$ for $k>1$.

For transient walks ($\mu>1$), both  the number of visited sites and the number of sites on the interface between visited and non-visited site grow linearly with time \cite{Mariz2001}. Then Eq.~(2) in the main text gives $P_\text{trap}(\tau)$ is a constant t, and we we denote this constant as $c$. Thus the early-time first-passage probability to a trap has the exponential form
\begin{align}
    F_n(\tau)=(1-c)^{\tau-1}c \; .
\end{align}
If this exponential regime breaks down at a time $t_n=\left( \ln n \right)^{1/\Delta}$ which diverges as  $n\to \infty$, then we have 
\begin{align}
\left\langle \tau_n^k \right\rangle \sim \sum_{\tau=1}^{t_n} \tau^k F_n(\tau) \sim \sum_{\tau=1}^{\infty} \tau^k (1-c)^{\tau-1}c=\text{const}\;.
\end{align}

\subsubsection{Numerical check of the moments }
Here we present the results of the numerical simulations and verify the predicted moments obtained in the previous paragraph.
\begin{figure}[th!]
    \centering
    \includegraphics[width=\columnwidth]{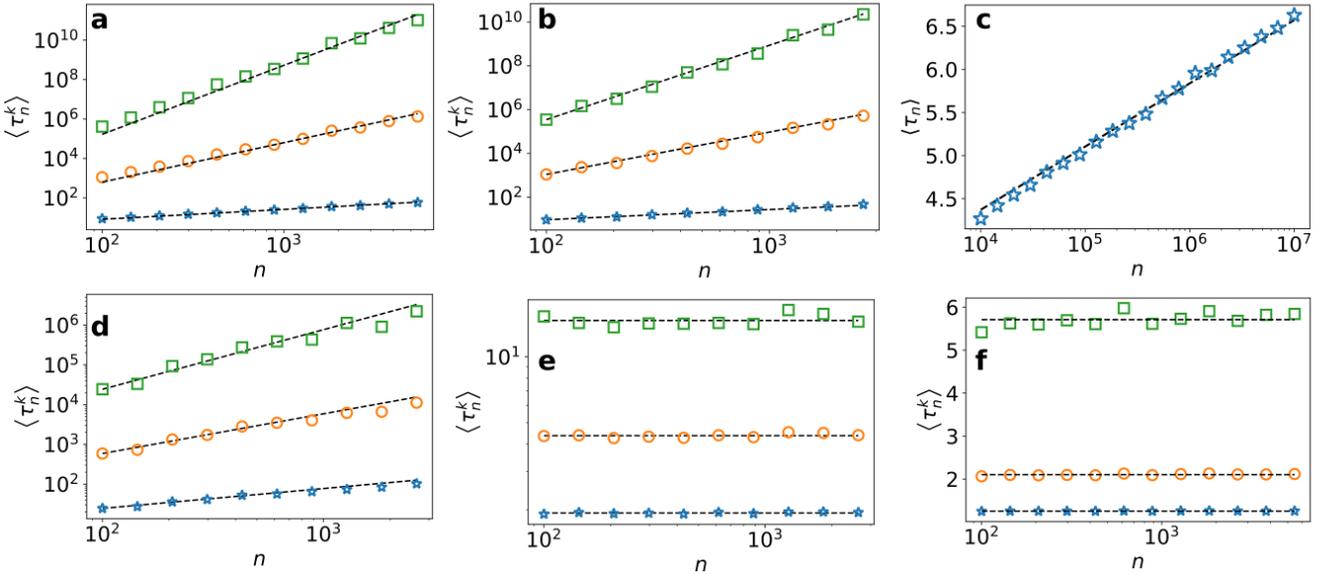}
     \caption{{\bf Moments of $\tau_n$:} the predicted scaling behaviours are in black dashed line. We show the examples of {\bf a} $1d$ L\'evy flights of parameter $\alpha=1.5$ , {\bf b} Sierpinski gasket nearest neighbour RW, {\bf d} $2d$ lattice nearest neighbour RW, {\bf e} $4d$ lattice nearest neighbour RW, {\bf f} $1d$ L\'evy flights of parameter $\alpha=0.5$. {\bf c} is the first moment of $\tau_n$ for the $2d$ lattice compared to $\frac{1}{\pi}\ln (n)$. For all figures except {\bf d}, we represent the moments $\left\langle \tau_n^k \right\rangle$ for $k=1$ (blue stars), $k=2$ (orange circles) and $k=3$ (green squares). For {\bf d},  blue stars are related to $k=2$, orange circles to $k=3$, and green squares to $k=4$.}
    \label{fig:Moments}
\end{figure}

\section{Numerical simulations}

\subsection{Recurrent random walks} \label{sec:WalkModel}

\begin{figure}[th!]
\includegraphics[width=\columnwidth]{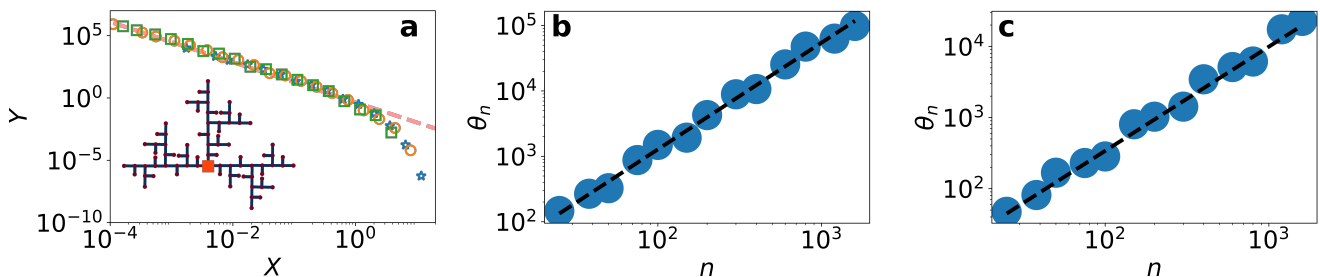}
\caption{{\bf Recurrent random walks ($\mu<1$):} {\bf a} Universal distribution of the time between visits to new sites for a 9-generation T-tree for nearest-neighbor random walks. Plotted are simulation results for the scaled distribution $Y\equiv \theta_n^{1+\mu}F_n(\tau)$ versus $X\equiv\tau/\theta_n$ for $n=100,\;500$, and $1000$ (respectively blue stars, orange circles, and green squares), where $\theta_n$ is the decay time in the exponential tail: $F_n(\tau)\sim \exp\left(-\tau/\theta_n\right)$. The dashed line indicates the algebraic decay $A(\mu)t^{-1-\mu}$ (see section S3.A.{\it4} for the definition of $A(\mu)$). The inset shows a 4-generation T-tree. {\bf b} $\theta_n$ as a function of $n$ for the T-tree compared to $n^{1/\mu}$ in black dashed line ($\mu=\frac{\ln 3}{\ln 6}$). {\bf c} $\theta_n$ as a function of $n$ for the Sierpinski gasket compared to $n^{1/\mu}$ in black dashed line ($\mu=\frac{\ln 3}{\ln 5}$).}\label{fig:fractals}
\end{figure}

The recurrent case is illustrated by L\'evy flights in $1d$ with $\alpha\in]1,2[$ and by random walks on the Sierpinski gasket (fractal dimension $d_\text{f}=\log 3/\log 2$ and walk dimension $d_\text{w}=\log 5/\log 2$), the T-tree (\ref{fig:fractals}{\bf a}) with $d_\text{f}=\log 3/\log 2$ and $d_\text{w}=\log 6/\log 2$), and $2d$ critical percolation clusters ($d_\text{f}=91/48$ and $d_\text{w}\approx 2.878$), using the algorithms described in Ref.~\cite{wolfram2002}. The Sierpinski gasket in our simulations is unbounded, while the T-tree has 9 generations. For these two examples, all walks start at the central site (inset to Fig.~3{\bf b} of the main text for the Sierpinski gasket and \ref{fig:fractals}{\bf a} for the T-tree). The critical percolation cluster was constructed from a $1000\times1000$ periodic square lattice, from which half of the bonds were randomly removed and then the largest cluster was selected. Simulations were performed on the $31$ clusters whose size is closest to the median size (out of an ensemble of $1000$ clusters), with the starting site chosen randomly. L\'evy flights are defined as follows: a jump length is drawn from the distribution $p(s)=1/[2\zeta(1+\alpha)|s|^{1+\alpha}]$ for $s = \pm 1 , \; \pm 2 , \; ...$. Each jump takes one time unit. Intermediate sites between the starting and final sites of each jump are not visited. 
All these examples support the properties of the distribution $F_n(\tau)$ for $\mu<1$ summarized in Table I of the main text, see also Fig.~3 of the main text and \ref{fig:fractals}\textbf{a} for the T-tree.

In \ref{fig:fractals}{\bf b} and {\bf c}, we show that $\theta_n$ the decay rate of the exponential regime indeed scales as $n^{1/\mu}$ in the examples of the T-tree and the Sierpinski gasket. We emphasize that the scaling of the moments shown in \ref{fig:Moments} and the scaling form of the distribution shown in Fig.~3 of the main text with $\theta_n$ are another direct demonstration of the existence of a single time scale $t_n=T_n=n^{1/\mu}$. 
\subsection{Exact enumeration technique} \label{sec:ExEnum}

Evaluating the large time statistics $F_n(\tau)$ of $\tau_n$, which can be of order $<10^{-10}$ (see Fig.~3 of the main text), cannot be done by direct sampling of the event $\lbrace \tau_n=t \rbrace$, both for the recurrent and the transient cases. To overcome this problem, we rely on an exact enumeration technique adapted from Ref.~\cite{Majid1984}, which consists in evaluating the exit time statistics of the visited domain without generating the exiting trajectory but only the visited domain. The transient case also requires the Wang-Landau procedure~\cite{Wang_Landau_PRL,Wang_Landau} and importance sampling~\cite{tokdar2010importance}, as described in section \ref{sec:MCMC}, because rare configurations of visited domains take part in the time statistics, contrary to the recurrent case (see Sec.~\ref{sec:stretched}). 

We now explain our method in detail. Consider the situation when $n$ distinct sites have been visited. We start from the last visited site. We are interested in the first exit time of the visited domain. We define $\mathbf{r}_n$ as the vector whose entries are given by the $n$ distinct sites successively visited. Then we define $\mathbb{P} \left(\tau_n=t|\mathbf{r}_n \right)$ as the first exit-time probability of the RW at time $t$ from the list of sites $\mathbf{r}_n$. We build a transition matrix $M(\mathbf{r}_n)$ from $\mathbf{r}_n$ based on the single-step transition probability from the $i^\text{th}$ visited site to the $j^\text{th}$ visited site, $p(i\to j)$,
\begin{align}
 \label{eq:trans_mat}
 M(\mathbf{r}_n)_{j,i}\equiv p\left(i \to j \right) \; .
\end{align}
Next, we define $v$ as the vector of length $n$ with entries $0$ except for the $n^\text{th}$ one which is $1$ ($v_{i_0}=\delta_{i_0,n}$). As a consequence of the definition \eqref{eq:trans_mat}, and taking $t\geq 0$, we have that
\begin{align}
    \sum_i\left(M(\mathbf{r}_n)^t \cdot v \right)_i &=\sum_{i_0,\ldots,i_{t-1},i}M(\mathbf{r}_n)_{i,i_{t-1}} \ldots M(\mathbf{r}_n)_{i_1,i_0} v_{i_0} \nonumber\\
    &=\sum_{i_1,\ldots,i_{t-1},i}p(n \to i_1) p(i_1 \to i_2) \ldots p(i_{t-1} \to i)\label{eq:exact_enum} \; .
\end{align}
We note that $p(n \to i_1) p(i_1 \to i_2) \ldots p(i_{t-1} \to i)$ is the probability to be successively at the $i_1^\text{th}$, ..., $i_{t-1}^\text{th}$, $i^\text{th}$ visited site
when starting from the $n^\text{th}$ visited site. Thus Eq.~\eqref{eq:exact_enum} expresses the exact enumeration of all of the trajectories of length $t$ going through visited sites \emph{only}, starting from the last visited site, by summing over all the values of $i_1$, ..., $i_{t-1}$, $i$. As a consequence, Eq.~\eqref{eq:exact_enum} is simply the probability of exiting the visited domain at a time larger or equal to $t+1$, $\mathbb{P}(\tau_n \geq t+1|\mathbf{r}_n)$. This results in
\begin{equation}
\mathbb{P}\left(\tau_n=t|\mathbf{r}_n\right)=\sum\limits_{i=1}^n \left(\left(M(\mathbf{r}_n)^{t-1}-M(\mathbf{r}_n)^t\right) \cdot v\right)_i \; .\label{eq:Ptaun}
\end{equation}
Now we use the following relation
\begin{align}
\mathbb{P}\left(\tau_n=t\right)=\sum\limits_{\mathbf{r}_n}\mathbb{P} \left(\tau_n=t|\mathbf{r}_n \right) \widetilde{\Pi} \left(\mathbf{r}_n \right)\; ,
\end{align}  
where $\widetilde{\Pi} (\mathbf{r}_n)$ is the generation probability to obtain the visited domain $\mathbf{r}_n$ of size $n$. Thus, to compute $\mathbb{P}\left(\tau_n=t\right)$, we are left with the sampling of $\mathbf{r}_n$, and the evaluation of $\mathbb{P} \left(\tau_n=t|\mathbf{r}_n \right)$ by Eq.~\eqref{eq:Ptaun}. For recurrent and marginal RWs, $\mathbf{r}_n$ is sampled simply by drawing successive steps until $n$ sites have been visited. For persistent random walks on hypercubic lattices, $M(\mathbf{r}_n)$ is now a $2dn \times 2dn$ matrix, where one adds to the description of a state its previous direction (there are $2d$ different directions).

\subsection{Monte Carlo simulations of transient RWs}
\label{sec:MCMC}

As explained in Sec.~\ref{sec:ExEnum}, for the transient RWs, we need, in addition to the exact enumeration, to sample the rare states of large trap-free regions. To do so, we use the technique of importance sampling \cite{tokdar2010importance} and the method developed by Wang and Landau \cite{Wang_Landau_PRL,Wang_Landau}. 

We focus on nearest-neighbor RWs on hypercubic lattices. Here $\boldsymbol{\sigma}_n = \boldsymbol{\sigma}$ ($n$ implicit) is defined by the list of the $M_{\boldsymbol{\sigma}}$ successive steps $\sigma_1,...,\; \sigma_{M_{\boldsymbol{\sigma}}}$ performed by the RW before having visited $n$ distinct sites. This additional information is required for the Monte Carlo algorithm, as now the new generation probability $\Pi(\boldsymbol{\sigma})$ is given by
\begin{align}
\Pi(\boldsymbol{\sigma})=p(\sigma_1)\ldots p(\sigma_{M_{\boldsymbol{\sigma}}}) \; .
\end{align}
One can easily go from $\boldsymbol{\sigma}$ to $\mathbf{r}_n$ using that $\mathbf{r}_n$ can be obtained from the list $\big[ 0, \sigma_1, \sigma_1+\sigma_2, \ldots, \sum\limits_{i \geq M_{\boldsymbol{\sigma}}} \sigma_i\big]$ by removing sites whenever they reappear, keeping the first occurrence at the same position. Thus, we define $\mathbb{P} \left(\tau_n=t|\boldsymbol{\sigma} \right)$ as $\mathbb{P} \left(\tau_n=t|\mathbf{r}_n \right)$ with the list $\mathbf{r}_n$ related to $\boldsymbol{\sigma}$.

\subsubsection{Importance sampling: general idea}

The idea is to bias the generation probability $\Pi(\boldsymbol{\sigma})$ towards states which contribute the most to the statistics we want to sample. Here, the statistics to sample is $\mathbb{P} \left( \tau_n=t \right)$, which can be seen as the average of the observable $\mathbb{P}\left(\tau_n=t|\boldsymbol{\sigma} \right)$ over the generation probability as
\begin{align}
\mathbb{P} \left( \tau_n=t \right) = \sum\limits_{\boldsymbol{\sigma}}\mathbb{P}\left(\tau_n=t|\boldsymbol{\sigma}\right)\Pi
(\boldsymbol{\sigma}) \; .
\end{align}
We associate to every state $\boldsymbol{\sigma}$ an energy $E(\boldsymbol{\sigma})$, such that rare states are associated to low energies. We now bias the generation probability $\Pi(\boldsymbol{\sigma})$ in the following way,
\begin{align}
\label{ImpSamp}
\mathbb{P} \left( \tau_n=t \right) = \sum\limits_{\boldsymbol{\sigma}}\mathbb{P}\left(\tau_n=t|\boldsymbol{\sigma} \right)\Pi
(\boldsymbol{\sigma})
    = \sum\limits_{\boldsymbol{\sigma}} \mathbb{P}\left(\tau_n=t|\boldsymbol{\sigma} \right)\frac{g\left(E(\boldsymbol{\sigma}) \right)}{ g\left(E(\boldsymbol{\sigma}) \right)}\Pi(\boldsymbol{\sigma}) \; , 
\end{align}
where $g$ is the bias that we want to introduce. This allows us to define the biased probability distribution $\Pi_g(\boldsymbol{\sigma}) = \Pi(\boldsymbol{\sigma}) g\left(E(\boldsymbol{\sigma}) \right)^{-1}Z_g^{-1} $, where $Z_g$ is a normalization constant.  Thus we have
\begin{align}
 \mathbb{P} \left( \tau_n=t \right)  = \sum\limits_{\boldsymbol{\sigma}}\mathbb{P}\left(\tau_n=t|\boldsymbol{\sigma} \right)  g\left(E(\boldsymbol{\sigma}) \right)Z_g\Pi_g(\boldsymbol{\sigma})
   = \frac{ \left\langle  \mathbb{P}\left(\tau_n=t|\boldsymbol{\sigma}\right) g\left(E(\boldsymbol{\sigma}) \right)\right\rangle_g }{\left\langle g\left(E(\boldsymbol{\sigma}) \right)\right\rangle_g},
\end{align}
where we have used $Z_g^{-1}= \left\langle g\left(E(\boldsymbol{\sigma}) \right) \right\rangle_g$, since $\sum\limits_t \mathbb{P} \left( \tau_n=t  | \boldsymbol{\sigma} \right)=1$, and noting $\left\langle \ldots \right\rangle_g$ the average with respect to the distribution $\Pi_g$. The terms $\mathbb{P}\left(\tau_n=t|\boldsymbol{\sigma}\right)$ are computed as described in section \ref{sec:ExEnum}. If the bias is wisely chosen (by enhancing the probability of low-energy states) the error in the estimation of the average of $g\left(E(\boldsymbol{\sigma}) \right) \mathbb{P}\left(\tau_n=t|\boldsymbol{\sigma} \right) Z_g^{-1} $, drawn from $\Pi_g(\boldsymbol{\sigma})$, will be smaller than the error that comes from $\mathbb{P}\left(\tau_n=t|\boldsymbol{\sigma} \right)$, drawn from $\Pi(\boldsymbol{\sigma})$. Thus, this technique allows us to sample statistics that are usually not observable, because of the error made by averaging is much bigger than the estimated value itself.

\subsubsection{Choice of the energy}

In this section, we choose a definition of the energy $E(\boldsymbol{\sigma})$, and give the reasons that motivate this choice. We start from Eq.~\eqref{eq:Kac}: the (average) return time to the boundary (where traps are located) is inversely proportional to the domain size, {\it i.e.,} the perimeter $P$ of the visited domain of fixed volume $n$. Thus, small perimeters $P$ correspond to the large exit times statistics we want to sample. Additionally, defining $L$ as the number of pairs $(i,j)$ of visited sites at distance 1 (with no regards on the order), we have the following relation between $P$ and $L$
\begin{align}
P+2L=2dn
\end{align}
for a $d$-dimensional hypercubic lattice. Indeed, each of the $n$ sites has $2d$ nearest neighbors (visited or not). Thus, one needs to maximize $L$ in order to minimize $P$, which in turn gives access to the exit-time statistics at long times. These considerations result in the following choice for the energy of a state $\boldsymbol{\sigma}$:
\begin{equation}
E(\boldsymbol{\sigma}) \equiv -L \; .
\end{equation}

\subsubsection{The Monte Carlo moves: example}
One needs to define the steps that can be made between successive states $\boldsymbol{\sigma}$. Similar to \cite{Barkema_2001}, we developed the following algorithm, based on successive step exchanges:

\begin{algorithm}
\caption{Monte-Carlo move algorithm, MC($\boldsymbol{\sigma}$)}
Given a state $\boldsymbol{\sigma}$ of steps $\sigma_1 \;  \ldots,\; \sigma_{M_{\boldsymbol{\sigma}}}$ of $n$ distinct visited sites \; 
Choose uniformly one index $i$ between $1$ and $M_{\boldsymbol{\sigma}}$ (included) \;
    \eIf{ $i<M_{\boldsymbol{\sigma}}$}{ 
    Exchange steps $\sigma_i$ and $\sigma_{i+1}$  
    }
   {
         Draw a new step $\sigma_{M_{\boldsymbol{\sigma}}+1}$ from $p$ \;
         Add it to the list of steps \;
         Exchange steps $\sigma_{M_{\boldsymbol{\sigma}}}$ and $\sigma_{M_{\boldsymbol{\sigma}}+1}$ \;
         }
    \While{ \text{the number of visited sites $<n$}}{ 
    Add steps drawn from the transition probability $p$ to the end of the list of steps \; 
    }
    \While{\text{ the number of visited sites $>n$ or the last step points to a site that has been visited several times}}{ 
    Remove the last step \; 
    }
    Return the new state $\boldsymbol{\sigma}'$
\end{algorithm}

\begin{figure}[ht!]
\includegraphics[width=0.6\columnwidth]{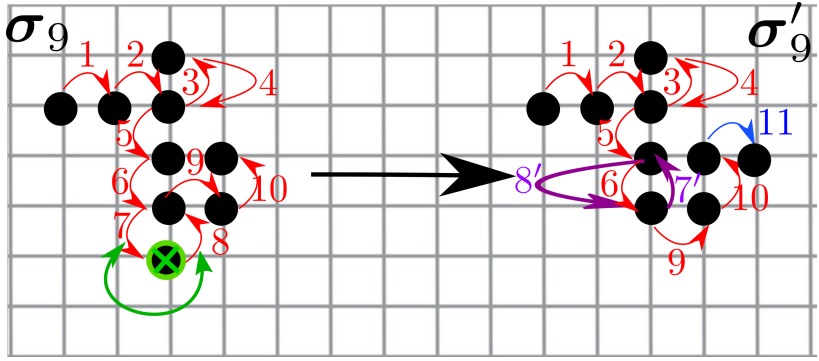}
\caption{{\bf Example of Monte Carlo move.} Steps are red arrows, visited sites are in black. The suppressed visited site is green crossed. The green arrow points to the the two steps (in purple) that have been exchanged, and in blue the added step to keep the number of distinct sites visited fixed. }
\label{fig:lattice_MC}
\end{figure}
We illustrate the method for the example in \ref{fig:lattice_MC}. We start from a state $\boldsymbol{\sigma}_9$ of $10$ steps (red arrows). By drawing uniformly between $1$ and $10$, we obtain $7$: we now exchange the $7^\text{th}$ and $8^\text{th}$ steps indicated by the green arrow. We obtain a new list of steps where $7'$ is the previous $8^\text{th}$ step, $8'$ is the previous $7^\text{th}$ step, and the rest of the trajectory is the same. However, this new list of steps gives one less visited site (green cross). Thus we add a new step $11$ (blue arrow) randomly drawn from the transition probability (here, because we make a nearest-neighbor RW, we choose uniformly between the 4 directions).  This new step leads to the visit of a new site. Thus we end up with a new list  $\boldsymbol{\sigma}'_9$, which consists of $11$ RW steps that correspond to $9$ distinct visited sites. 

This algorithm results in the following probability of proposing a given state $\boldsymbol{\sigma}'\equiv\boldsymbol{b}$ ($M_{\boldsymbol{b}}$ steps) starting from state $\boldsymbol{\sigma}\equiv\boldsymbol{a}$ ($M_{\boldsymbol{a}}$ steps),
\begin{align}
\label{MCacc}
    p_{\text{prop}}(\boldsymbol{a} \to \boldsymbol{b}) = \left\{
    \begin{array}{ll}
      {\displaystyle   \frac{p(b_{M_{\boldsymbol{a}}+1})\ldots p(b_{M_{\boldsymbol{b}}})}{M_{\boldsymbol{a}}}}&\mbox{if $M_{\boldsymbol{b}}>M_{\boldsymbol{a}}$,} \\
      {\displaystyle  \frac{1}{M_{\boldsymbol{a}}}} & \mbox{otherwise,} 
    \end{array} \right.
\end{align}
given that $\boldsymbol{b}$ can be obtained from $\boldsymbol{a}$ with the successive steps exchange algorithm. The factor $1/M_{\boldsymbol{a}}$ corresponds to the choice of the step we change (either exchange of step $i$ and $i+1$ or change of the last one if $i=M_{\boldsymbol{a}}$), $p(b_i)$ being the transition probability to make the jump $b_i$ in one time step ($p(b_i)=1/(2d)$ for the nearest-neighbor RW on the hypercubic lattice). The other factors correspond to the right choice of successive added steps which lead to $\boldsymbol{b}$. 

Now that we have our algorithm to move from one state to another, we return to the problem of sampling $\Pi_g$. To this end, we consider the detailed balance condition
\begin{align}
\label{eq:det_bal}
    \pi(\boldsymbol{a} \to \boldsymbol{b})\Pi_g(\boldsymbol{a})=\pi(\boldsymbol{b} \to \boldsymbol{a})\Pi_g(\boldsymbol{b}) \; , \qquad
    \text{with }\quad \pi(\boldsymbol{a}\to \boldsymbol{b})=p_{\text{prop}}(\boldsymbol{a} \to \boldsymbol{b})p_{\text{acc}}(\boldsymbol{a} \to \boldsymbol{b}) \; ,
\end{align}
where $p_{\text{acc}}(\boldsymbol{a} \to \boldsymbol{b})$ is the probability to accept the Monte Carlo move. We use a classical result from Markov chain theory: if we use an algorithm able to draw transitions between states with probability $\pi (\boldsymbol{a} \to \boldsymbol{b})$ following the detailed balance condition of Eq.~\eqref{eq:det_bal}, then the states obtained from the algorithm are distributed according to $\Pi_g$ in the steady state.
Now, using that 
\begin{align}
    \Pi_g(\boldsymbol{a})=p(a_1)\ldots p(a_{M_{\boldsymbol{a}}}) g(E(\boldsymbol{a}))^{-1}Z_g^{-1} \; ,
\end{align}
the detailed balance condition becomes using Eqs.~\eqref{MCacc} and \eqref{eq:det_bal}, and taking without loss of generality $M_{\boldsymbol{a}}\leq M_{\boldsymbol{b} }$,
\begin{equation}
    p_\text{acc}(\boldsymbol{a} \to \boldsymbol{b})g(E(\boldsymbol{a}))^{-1}Z_g^{-1}\frac{1}{M_{\boldsymbol{a}}}p(a_1) \ldots p(a_{M_{\boldsymbol{a}}})p(b_{M_{\boldsymbol{a}}+1})\ldots p(b_{M_{\boldsymbol{b}}}) =p_\text{acc}(\boldsymbol{b} \to \ \boldsymbol{a})g(E(\boldsymbol{b}))^{-1}Z_g^{-1}\frac{1}{M_{\boldsymbol{b}}}p(b_1) \ldots p(b_{M_{\boldsymbol{b}}})
    \label{eq:MCcond}
\end{equation}
in the case for which the algorithm permits a move from $\boldsymbol{a}$ to $\boldsymbol{b}$ in a single Monte Carlo step. Thus we take the following acceptance rate using the Metropolis-Hastings prescription \cite{metropolis1953equation}
\begin{equation}
p_{\text{acc}}(\boldsymbol{a} \to \boldsymbol{b}) =  \min \left( 1,\frac{M_{\boldsymbol{a}} g(E(\boldsymbol{a}))}{M_{\boldsymbol{b}} g(E(\boldsymbol{b})) }  \right) \; .
\label{eq:acc_fin}
\end{equation}
%\clearpage
\subsubsection{Choice of the bias: Wang-Landau}
To use the procedure described in the previous subsection we have to make a choice of the bias $g$. We follow the Wang-Landau prescription to sample uniformly the energy levels, biasing them via the approximation of the energy distribution obtained by the algorithm \cite{Wang_Landau_PRL,Wang_Landau}. To obtain $g(E)$, the idea is to explore the energy landscape $E(\boldsymbol{\sigma})$ via the Monte-Carlo moves described in Sec.~\ref{sec:MCMC}. Every time we observe an energy $E$, we update our knowledge of the energy density $g(E)$ and use Eq.~\eqref{eq:acc_fin} to bias the random walk on the energy landscape towards rare states (low values of $g(E)$). This has the effect of flattening the observed energy landscape, as can be seen by computing the energy histogram $H(E)$. We end the algorithm when we observe each energy level with similar probability, {\it i.e.}, when the energy landscape is flat enough.

A realization of this algorithm is the following: 
\begin{algorithm}
\caption{Wang-Landau algorithm to obtain the energy density $\mathcal{N}(E)$}
Given a state $\boldsymbol{\sigma}$, a uniform distribution $g(E)$, a factor $f=\exp (1)$ and an energy histogram $H(E)$ with entries $0$
\While{$f>\exp(10^{-8})$}{
        Draw $\boldsymbol{\sigma}'$ from MC($\boldsymbol{\sigma}$) \;
        Replace $\boldsymbol{\sigma}$ by $\boldsymbol{\sigma}'$ with probability $p_\text{acc}(\boldsymbol{\sigma} \to \boldsymbol{\sigma}')$ (see Eq.~\eqref{eq:acc_fin}) \;
        Multiply $g(E(\boldsymbol{\sigma}))$ by $f$, $g(E(\boldsymbol{\sigma})) \to f \times g(E(\boldsymbol{\sigma}))$ \; 
        Increase the energy histogram $H(E(\boldsymbol{\sigma}))$ by 1, $H(E(\boldsymbol{\sigma}))\to 1+ H(E(\boldsymbol{\sigma}))$ \; 
        \If{ $\sum\limits_E H(E)>1000$ and each value of $H(E)$ is larger than $0.9 \frac{1}{N_E}\sum\limits_E H(E)$ ($N_E$ the number of energy levels)}
        {
         Reset all the values of the histogram $H(E)$ to 0 \; 
         Replace the value of $f$ by $\sqrt{f}$ \;
        }
    }
    Return $g(E)/\sum\limits_E g(E)$ which approximates the energy density $\mathcal{N}(E)$
\end{algorithm}

\subsubsection{Results of the Monte Carlo simulations}
\begin{figure}[th!]
    \centering
\includegraphics[width=\columnwidth]{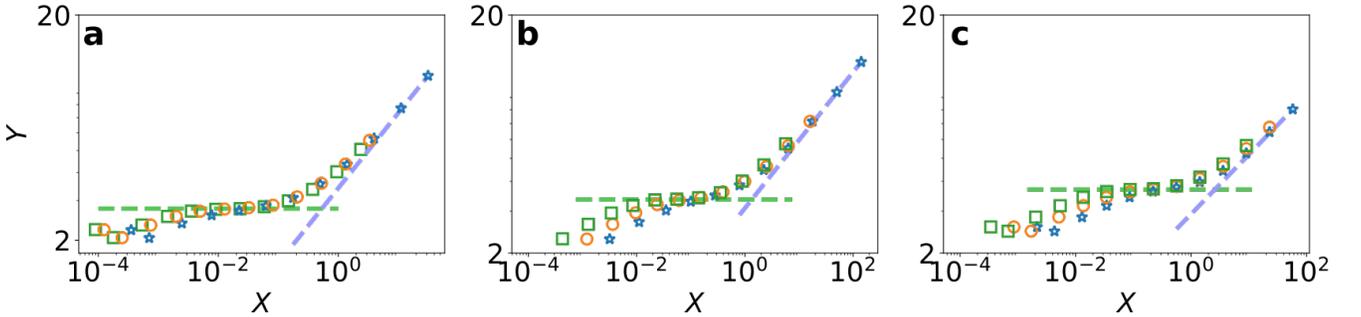}
    \caption{{\bf Transient nearest neighbour RWs $(\mu>1)$ on hypercubic lattice.} The scaled distribution $Y\equiv \left[-(\ln t_n)^2 F_n(\tau)\right]/(\tau/t_n)^{\mu/(1+\mu)}$ versus $X\equiv\tau/T_n$ for nearest-neighbor RWs in: {\bf a} $d=4$ (for $n=200$, $400$, and $500$, blue stars, orange circles, and green squares, respectively), {\bf b} $d=5$, and {\bf c} $d=6$ (for $n=100$, $200$, and $400$, blue stars, orange squares, and green squares, respectively). The early-time plateau illustrates the stretched exponential regime (horizontal green dashed line) and the blue dashed line, proportional to $X^{1/(1+\mu)}$, corresponds to the exponential regime.}
    \label{fig:Trans}
\end{figure}

We apply the method developed above to nearest-neighbor RWs on the cubic lattice and obtained Fig.~4{\bf c} of the main text. The results for hypercubic lattices in dimensions $d=4,\; 5$ and $6$ are displayed in \ref{fig:Trans}. Using appropriately scaled units, we verify the stretched exponential and the exponential long-time regimes.

\subsubsection{Stretched exponential without the Monte Carlo method }
\begin{figure}[th!]
    \centering
\includegraphics[width=0.8\columnwidth]{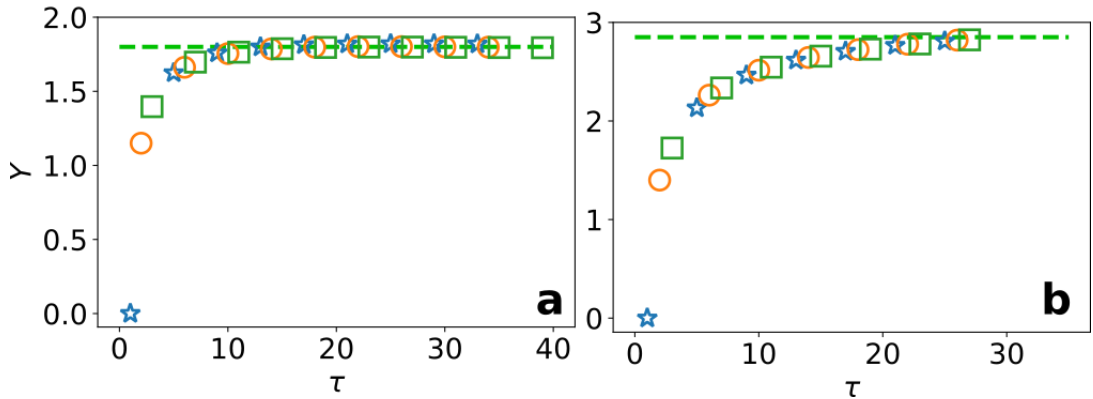}
    \caption{{\bf Transient L\'evy flights:} $Y=\frac{-\ln F_n(\tau)}{\tau^{\mu/(1+\mu)}}$ for L\'evy flights of parameter {\bf a} $\alpha=0.5$ in $1d$ {\bf b} $\alpha=1.2$ in $3d$ for $n=400$, $800$, and $1600$ (blue stars, orange circles, and green squares, respectively). Both are compared to a constant value (dashed line).}
    \label{fig:TransLevy}
\end{figure}

Even for random walks where the Monte Carlo technique of Sec S4.C.{\it5} does not apply, we can still observe the beginning of the stretched exponential law using the exact enumeration method. Here we show in \ref{fig:TransLevy} two other transient walks, the $1d$ L\'evy flight of parameter $\alpha=0.5$ and the $3d$ L\'evy flight of parameter $\alpha=1.2$. Both display the stretched exponential behaviour as expected.
\section{Extensions}
\subsection{General observables} \label{sec:motif}
\subsubsection{Definition of bulk and boundary observables}

A more complete characterization of the visitation statistics of random walks is provided by introducing two classes of more general visitation observables. A visitation observable is called {\em bulk} if only visited sites are involved in its characterization. Otherwise it is a {\it boundary} (or surface) observable (see \cite{van1996topology,Mariz2001} for precise definitions). According to these definitions, the number $L(t)$ of links/adjacent visited sites is a bulk observable (illustrated in \ref{fig:Motifs}), while the perimeter $P(t)$ of the visited domain and the number $I(t)$ of islands are boundary observables.

\subsubsection{Dynamics of bulk observables}

\begin{figure}[th!]
    \centering
 \includegraphics[width=\columnwidth]{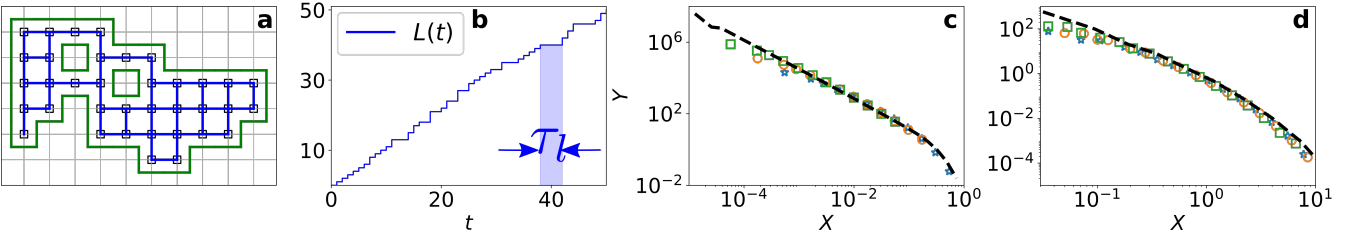}
    \caption{{\bf Extension to general "bulk" observables for recurrent and marginal RWs.} {\bf a} An example of bulk observable: the number of links in the support of the walk. The blue lines are the links in the visited domain. {\bf b} Time dependence of the number of links $L(t)$. The shaded area represents the waiting time $\tau_l$ between two increments of $L(t)$. {\bf c} and {\bf d} Comparison between the rescaled distributions of $\tau_l$, $Y\equiv t_l^{1+\mu}F_l(\tau/t_l)$ versus $X=\tau/t_l$ (color) and the rescaled distributions of $\tau_n$, $Y\equiv t_n^{1+\mu}F_n(\tau/t_n)$ versus $X\equiv \tau/t_n$ (black) for {\bf c} nearest-neighbor RWs on the square lattice and {\bf d} L\'evy flights of parameter $\alpha=1/\mu=\ln 6/\ln 3$.  In these plots, $l$ takes the values $100$, $200$ and $400$ (blue stars, orange circles, green squares), while $n=1000$ (black dashed line). Here $t_l=l^{1/\mu}$ if $\mu<1$, $t_l=\sqrt{l}$ if $\mu=1$, and similarly for $t_n$. }
    \label{fig:Motifs}
\end{figure}

We argue that the dynamics of bulk variables is the same as that for the number of new sites visited. For the example of the number of new links $L(t)$ visited, both for a nearest-neighbor RW on a square lattice and a $1d$ L\'evy flight with $\alpha=1$, we check in  \ref{fig:Motifs}{\bf c} and {\bf d} that the distribution of times $\tau_l$ between successive creation of links is similar to the distribution of times $\tau_n$ between successive visits to new sites.

\subsubsection{Dynamics of boundary observables}
\label{sec:exponentsurface}

We now derive the exponent $\phi$ of the algebraic decay $F_\Sigma (\tau)\sim \tau^{-1-\phi}$ for recurrent and marginal RWs for a boundary observable $\Sigma$, as defined in the main text. This derivation follows the spirit of Section S3.A.{\it 3}. The first step consists in realizing that the finite-size cutoff time $t_\Sigma$ in the dynamics of a surface observable $\Sigma$ is the same as the cutoff time $t_n\sim n^{1/\mu}$ for the number $n$ of distinct sites visited. Next, we use the known result (see \cite{van1996topology,Mariz2001}) that for a boundary observable $\Sigma$,
\begin{align}
\left\langle \Sigma(t) \right\rangle \sim t^{2\mu-1} \label{eq:sigma_t} \;.
\end{align}
Using the scaling of the number of distinct sites visited with time,  $t \sim n^{1/\mu}$, we obtain the cutoff time in terms of value of $\Sigma$ instead of $n$:
\begin{align}
t_\Sigma \sim \Sigma^{1/(2\mu-1)} \; .
\end{align}

We are now in a position to identify two expressions of the first moment. 
First, relying on its definition and the algebraic behavior $F_\Sigma(\tau)\propto \tau^{-1-\phi}$, we have
\begin{align}
\left\langle \tau_\Sigma \right\rangle &\propto \int_1^{t_\Sigma} \frac{t }{t^{1+\phi}}\;{\rm d}t \propto t_\Sigma^{1-\phi} = \Sigma^{(1-\phi)/(2\mu-1)} \;. \label{eq:tau_sig1}
\end{align}
Second, we use that
\begin{align}
\label{eq:der_sig}
\frac{{\rm d}\langle \Sigma \rangle}{{\rm d}t}\sim \frac{1}{\left\langle \tau_\Sigma \right\rangle}\sim t^{2\mu-2}\; ,
\end{align}
which leads, with the help of Eq.~\eqref{eq:sigma_t}, to
\begin{align}
\label{eq:tau_sig2}
\left\langle \tau_\Sigma \right\rangle \sim \Sigma^{(2 -2\mu)/(2\mu-1)}. 
\end{align}
Combining \eqref{eq:tau_sig1} and \eqref{eq:tau_sig2}, and matching the exponents, we finally obtain
\begin{equation}
    \phi=2\mu-1,
    \label{eq:expSurf}
\end{equation}
as given in the main text.

In the marginal case, we have \cite{van1996topology,Mariz2001}
\begin{align}
\left\langle \Sigma (t) \right\rangle &\sim \frac{t}{\ln^2 t}.
\end{align}
Using as previously in Eq.~\eqref{eq:der_sig} that
\begin{align}
 \frac{{\rm d} \left\langle \Sigma(t) \right\rangle}{{\rm d}t} & \sim \frac{1}{\left\langle \tau_{\left\langle \Sigma(t) \right\rangle} \right\rangle}\; ,
\end{align} 
we find 
\begin{equation}
   \langle \tau_\Sigma\rangle \sim \ln^2 t.  
\end{equation}
The definition of  $\langle \tau_\Sigma\rangle$ in terms of $F_\Sigma$
 finally yields
\begin{align}
    F_\Sigma(\tau)\propto \frac{\ln \tau}{\tau^2} \; ,
    \label{eq:SurfMarg}
\end{align}
as given in the main text. Equations \eqref{eq:expSurf} and \eqref{eq:SurfMarg} are numerically confirmed in Fig.~4{\bf b} and {\bf c} in the main text.

\subsection{Covariance of the number of distinct  sites visited for recurrent $1d$ L\'evy flights}
\subsubsection{Derivation of the multiple time correlations}
One can generalize the result on the covariance $\text{Cov}[N(t_1),N(t_2)]$, Eq. (6) of the main text, to 
\begin{align}
    \left\langle  (N(t_1)-\left\langle N(t_1) \right\rangle )\ldots (N(t_k)-\left\langle N(t_k)\right\rangle) \right\rangle \propto t_1^{1/\alpha}\ldots t_k^{1/\alpha}\; \frac{t_1}{t_k} \; .
    \label{eq:Covktimes}
\end{align}
To obtain this result, we start from (making the hypothesis that correlations between $\tau_k$ can be neglected)
\begin{align}
    \mathcal{L}\lbrace \mathbb{P}(N(t_1) \geq n_1, \ldots ,N(t_k) \geq n_k )\rbrace&=\frac{1}{s_1 \ldots s_k}\frac{h(0)}{h\big((s_1+\ldots+s_k)n_1^{1/\mu}\big)} \ldots \frac{h(s_kn_{k-1}^{1/\mu})}{h(s_kn_k^{1/\mu})}\; ,
    \label{eq:cov1}
\end{align}
where 
 \begin{align}
    h(s,n)&\equiv h(sn^{1/\mu})=\exp \left[  \int_0^n  \left(\mathcal{L}\lbrace \mathbb{P}(\tau_k=t) \rbrace -1\right) {\rm d}k \right] \; .
\end{align}
This is an extension of the calculation performed in \cite{Regnier2022}. By supposing that the correlations between $\tau_k$ can be neglected, one obtains a simple formula for the distribution of the number of distinct sites visited at $k$ different times as a product of functionals of the exit time distribution. We consider the limit $t_1 \ll \ldots  \ll t_k$, i.e., $s_1 \gg \ldots \gg s_k$ and $n_is_i^\mu=a_i$ fixed. For $k=2$, we notice that 
\begin{align}
&\mathcal{L}\lbrace \mathbb{P}(N(t_1) \geq n_1, N(t_2) \geq n_2 )\rbrace \nonumber \\
&\approx\frac{1}{s_1s_2}\frac{h(0)}{h(a_1^{1/\mu})}\frac{h(0)}{h(a_2^{1/\mu})} + \frac{s_2}{s_1}\frac{1}{s_1s_2}\left(\frac{h(0)}{h(a_1^{1/\mu})}\frac{h'(0)a_1^{1/\mu}}{h(a_2^{1/\mu})} -\frac{h(0)h'(a_1^{1/\mu})a_1^{1/\mu}}{h(a_1^{1/\mu})^2}\frac{h(0)}{h(a_2^{1/\mu})}\right)\nonumber \\
&= \mathcal{L}\lbrace \mathbb{P}(N(t_1) \geq n_1) \rbrace \mathcal{L}\lbrace \mathbb{P}(N(t_2) \geq n_2) \rbrace + a_1^{1/\mu}\frac{h'(0)h(a_1^{1/\mu})-h'(a_1^{1/\mu})}{s_1s_2 h(a_2^{1/\mu})h(a_1^{1/\mu})^2}\frac{s_2}{s_1} \; .
    \label{eq:covs}
\end{align}
Once integrated over $n_1$ and $n_2$, we obtain the covariance's scaling mentioned in the main text,
\begin{align}
    \text{Cov}\left[N(t_1),N(t_2) \right] \propto \frac{t_1}{t_2}t_1^{\mu}t_2^{\mu} \; .
    \label{eq:cov2times}
\end{align}
For $k=3$, we start from
\begin{align}
    &\left\langle (N(t_1)-\left\langle N(t_1)\right\rangle )(N(t_2)-\left\langle N(t_2)\right\rangle )(N(t_3)-\left\langle N(t_3)\right\rangle ) \right\rangle  \nonumber  \\ 
    =&\left\langle N(t_1)N(t_2)N(t_3)\right\rangle  -\left\langle N(t_1) \right\rangle\left\langle N(t_2)N(t_3)\right\rangle-\left\langle N(t_2) \right\rangle\left\langle N(t_1)N(t_3)\right\rangle \nonumber \\
    &-\left\langle N(t_3) \right\rangle \left\langle  N(t_1)N(t_2)\right\rangle +2\left\langle N(t_3) \right\rangle \left\langle  N(t_1)\right\rangle\left\langle N(t_2)\right\rangle \; .
\end{align}
Thus to compute the correlation function, we are left with the comparison of the three-time distribution which results in $\left\langle  N(t_1)N(t_2)N(t_3)\right\rangle$ and the sum of distributions giving rise to the other moments.
\begin{align}
    &\mathcal{L}\lbrace \mathbb{P}(N(t_1) \geq n_1,N(t_2) \geq n_2
    ,N(t_3) \geq n_3 )\rbrace=\mathcal{L}\lbrace \mathbb{P}(N(t_1) \geq n_1) \rbrace \mathcal{L}\lbrace \mathbb{P}(N(t_2) \geq n_2
    ,N(t_3) \geq n_3 )\rbrace \nonumber \\
    &+\mathcal{L}\lbrace \mathbb{P}(N(t_2) \geq n_2) \rbrace \mathcal{L}\lbrace \mathbb{P}(N(t_1) \geq n_1
    ,N(t_3) \geq n_3 )+\mathcal{L}\lbrace \mathbb{P}(N(t_3) \geq n_3) \rbrace \mathcal{L}\lbrace \mathbb{P}(N(t_1) \geq n_1
    ,N(t_2) \geq n_2 )\rbrace\rbrace \nonumber  \\
    &-2\mathcal{L}\lbrace \mathbb{P}(N(t_1) \geq n_1) \rbrace \mathcal{L}\lbrace \mathbb{P}(N(t_2) \geq n_2) \rbrace\mathcal{L}\lbrace \mathbb{P}(N(t_3) \geq n_3) \rbrace \nonumber \\
    &+a_1^{1/\mu}a_2^{1/\mu} \frac{ h'(0)^2h(a_1^{1/\mu})h(a_2^{1/\mu})-h'(0)h'(a_1^{1/\mu})h(a_2^{1/\mu})-h'(0)h(a_1^{1/\mu})h'(a_2^{1/\mu})+h'(a_1^{1/\mu})h'(a_2^{1/\mu}) }{s_1s_2s_3 h(a_3^{1/\mu})h(a_1^{1/\mu})^2h(a_2^{1/\mu})^2}\frac{s_3}{s_1}\nonumber \\
    &+o\left(\frac{s_3}{s_1} \right) \; . \label{eq:cov2}
\end{align}
By integrating this equation over $n_1$, $n_2$ and $n_3$ and using the Tauberian theorem  we obtain the scaling of \eqref{eq:Covktimes}. Generalising the calculation to any number of times gives \eqref{eq:Covktimes}.
\begin{figure}
    \centering
    \includegraphics[width=\columnwidth]{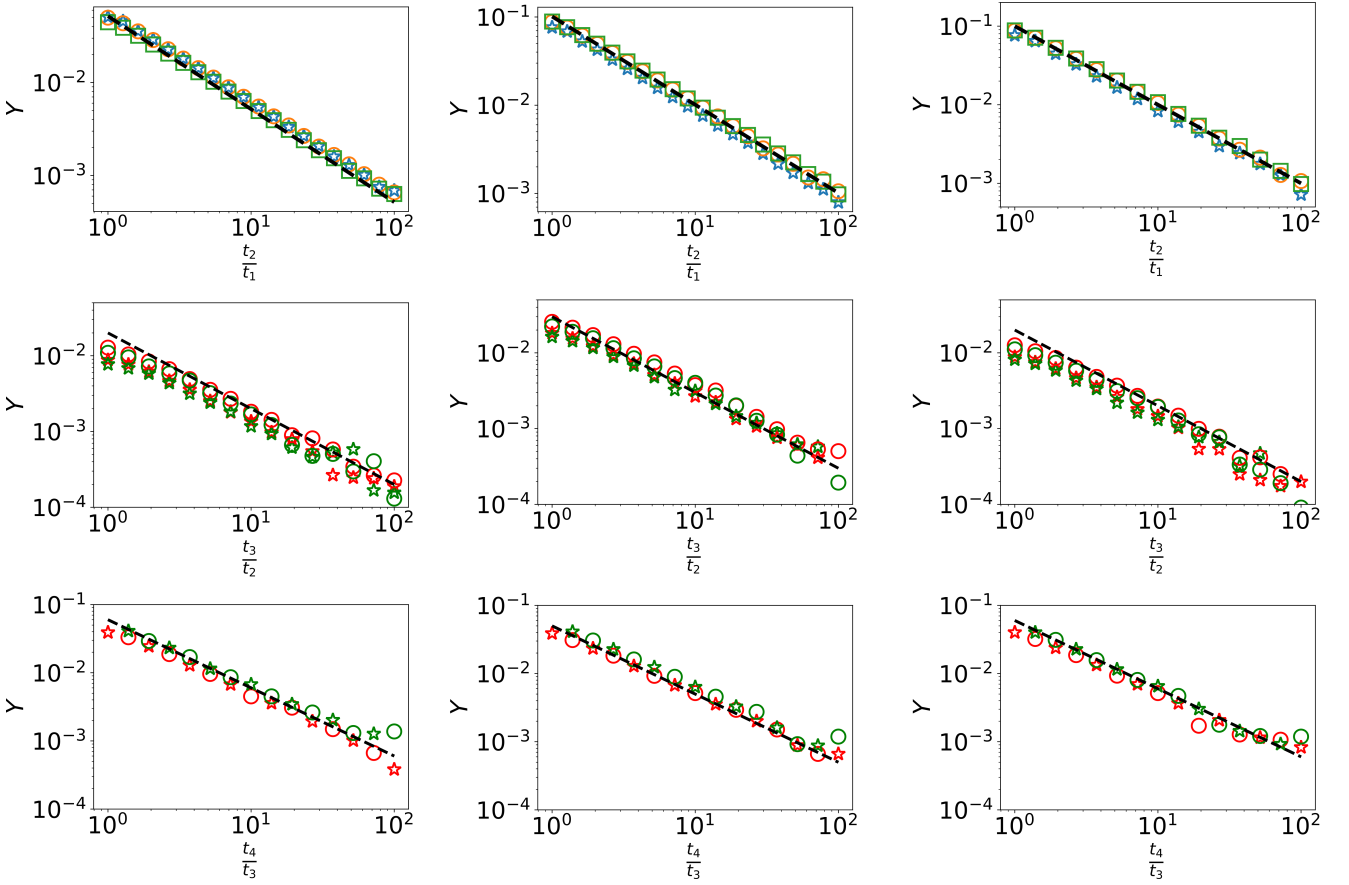}
    \caption{{\bf Multiple times correlation function of L\'evy flights:} The parameter $\alpha$ takes the values $\alpha=1.3$ (left column), $\alpha=1.5$ (middle column) and $\alpha=1.7$ (right column).\\
    Upper panel: $Y=\left\langle ((N(t_1)-\left\langle N(t_1)\right\rangle )(N(t_2)-\left\langle N(t_2)\right\rangle ))\right\rangle/(\left\langle N(t_1)\right\rangle \left\langle N(t_2)\right\rangle )  $, and we compare $Y$ to the black dashed line proportional to $\frac{t_1}{t_2}$. Stars stand for $t_1=10$, circles for $t_1=100$ and squares for $t_1=1000$. \\
    Middle panel: $Y=\left\langle((N(t_1)-\left\langle N(t_1)\right\rangle )(N(t_2)-\left\langle N(t_2)\right\rangle )(N(t_3)-\left\langle N(t_3)\right\rangle ))\right\rangle /(\left\langle N(t_1)\right\rangle\left\langle N(t_2)\right\rangle\left\langle N(t_3)\right\rangle)\frac{t_2}{t_1}$, and we compare $Y$ to the black dashed line proportional to $\frac{t_2}{t_3}$. Plots in red stand for $t_1=10$, in green $t_1=100$. Stars stand for $t_2=2t_1$ and circles for $t_2=4t_1$.\\
    Lower panel: $Y=\left\langle((N(t_1)-\left\langle N(t_1)\right\rangle )(N(t_2)-\left\langle N(t_2)\right\rangle )(N(t_3)-\left\langle N(t_3)\right\rangle )(N(t_4)-\left\langle N(t_4)\right\rangle ))\right\rangle /(\left\langle N(t_1)\right\rangle\left\langle N(t_2)\right\rangle\left\langle N(t_3)\right\rangle \left\langle N(t_4)\right\rangle)\frac{t_3}{t_1}$, and we compare $Y$ to the black dashed line proportional to $\frac{t_3}{t_4}$. Red symbols stand for $t_1=10$ and green for $t_1=100$. Stars stand for $t_2=2t_1$ and circles for $t_2=4t_1$. In all three lower subplots we take $t_3=4t_2$.}
    \label{fig:CovLevy}
\end{figure}
\subsubsection{Criterion on $\tau_n$ correlations}
Our starting point is a relation between the number of distinct sites visited and the sum of $\tau_k$,
\begin{align}
    \lbrace N(t) \geq n \rbrace =\left\lbrace \sum_{k=1}^{n-1}\tau_k \leq t  \right\rbrace \; .
\end{align}
Then, using that the rescaled process $N(t)/t^{\mu}$ is asymptotically independent of $t$ for recurrent random walks (rigorously proved in the mathematical literature in the case of recurrent L\'evy flights \cite{LeGall1991}), we have that $\sum_{k=1}^n\tau_k/n^{1/\mu}$ is independent of $n$ because
\begin{align}
    \mathbb{P}(N(t)/t^\mu \geq 1)=\mathbb{P}\left(\frac{\sum_{k=1}^{t^\mu-1}\tau_k}{t} \leq 1 \right) \; .
\end{align}
We note $\kappa_i \left( \sum_{k=1}^n\tau_k \right)=A_in^{i/\mu}$, $\kappa_i$ being the $i^{\rm th}$ cumulant.  We start from the relation between the number $N(t)$ of distinct sites visited and the $\tau_k$ and perform a cumulant expansion,
\begin{align}
    &\mathcal{L}\lbrace \mathbb{P}(N(t_1) \geq n_1,N(t_2) \geq n_2 )\rbrace\nonumber \\
    &=\sum_{t_1,t_2}\exp\left[-s_1t_1-s_2t_2\right]\mathbb{P}\left(\sum_{k=0}^{n_1-1}\tau_k \leq t_1, \sum_{k=0}^{n_2-1}\tau_k \leq t_2 \right) \nonumber \\
    &=\frac{1}{s_1s_2}\mathbb{E}\left(\exp\left[-s_1\sum_{k=0}^{n_1-1}\tau_k-s_2\sum_{k=0}^{n_2-1}\tau_k  \right] \right) \nonumber \\
    &= \frac{1}{s_1s_2}\exp\left[ \sum_{i=1}^\infty \frac{1}{i!}\kappa_i \left( -s_1\sum_{k=0}^{n_1-1}\tau_k-s_2\sum_{k=0}^{n_2-1}\tau_k \right)  \right]\; ,
\end{align}
which we compare to the product of single time distributions,
\begin{align}
    &\mathcal{L}\lbrace \mathbb{P}(N(t) \geq n )\rbrace=\frac{1}{s}\exp\left[ \sum_{i=1}^\infty \frac{1}{i!}\kappa_i \left( -s\sum_{k=0}^{n-1}\tau_k \right) \right]\equiv \frac{1}{s}H(sn^{1/\mu}) \; .
\end{align}
Keeping only the first non-zero terms (which appear with the second cumulant), we obtain
\begin{align}
    &\mathcal{L}\lbrace \mathbb{P}(N(t_1) \geq n_1,N(t_2) \geq n_2 )\rbrace-\mathcal{L}\lbrace \mathbb{P}(N(t_1) \geq n_1  )\rbrace \mathcal{L}\lbrace \mathbb{P}(N(t_2) \geq n_2  )\rbrace \nonumber \\
    &\approx \mathcal{L}\lbrace \mathbb{P}(N(t_1) \geq n_1  )\rbrace \mathcal{L}\lbrace \mathbb{P}(N(t_2) \geq n_2 )\rbrace \left(\exp\left[ s_1s_2 A_2 n_1^{2/\mu} +s_1s_2\text{Cov}\left[\sum_{k=0}^{n_1-1}\tau_k,\sum_{k=n_1}^{n_2-1}\tau_k \right]\right]-1 \right)\nonumber  \\
    &\approx \mathcal{L}\lbrace \mathbb{P}(N(t_1) \geq n_1  )\rbrace \mathcal{L}\lbrace \mathbb{P}(N(t_2) \geq n_2 )\rbrace \left( s_1s_2 A_2 n_1^{2/\mu} +s_1s_2\text{Cov}\left[\sum_{k=0}^{n_1-1}\tau_k,\sum_{k=n_1}^{n_2-1}\tau_k \right] \right) \; .
    \label{eq:devCov}
\end{align}
We are interested in the term which will dominate the covariance for values $t_1 \ll t_2$ i.e. $s_1 \gg s_2$. We integrate the first term over $n_1$ and $n_2$ to obtain the first term of the covariance, 
\begin{align}
    &\int {\rm d}n_1{\rm d}n_2 \mathcal{L}\lbrace \mathbb{P}(N(t_1) \geq n_1  )\rbrace \mathcal{L}\lbrace \mathbb{P}(N(t_2) \geq n_2 )\rbrace  s_1s_2 A_2 n_1^{2/\mu} \nonumber \\
    &\qquad \approx \frac{s_2}{s_1}\frac{1}{s_1^{1+\mu}s_2^{1+\mu}}\int {\rm d}a_1{\rm d}a_2   H(a_1^{1/\mu})H(a_2^{1/\mu})A_2 a_1^{2/\mu} \; .
\end{align}
We note that only the second term of \eqref{eq:devCov} contains the correlations between the $\tau_k$. Thus, the large time scaling of the covariance of the number of distinct sites visited is not affected by the correlations between the $\tau_k$ if the first term dominates the second one,
\begin{align}
     \ s_1s_2\;\text{Cov}\left[\sum_{k=1}^{n_1-1}\tau_k,\sum_{k=n_1}^{n_2-1}\tau_k \right]=O\left( s_1s_2 n_1^{2/\mu} \right)
\end{align}
or in other words, for any value of $n_2$
\begin{align}
     \text{Cov}\left[\sum_{k=0}^{n_1-1}\tau_k,\sum_{k=n_1}^{n_2-1}\tau_k \right]=O\left( n_1^{2/\mu} \right) \; .
     \label{eq:Criterion}
\end{align}
Thus, using the Tauberian theorem, under the condition 
%one has that if 
\eqref{eq:Criterion} 
%is verified
, we obtain \eqref{eq:cov2times}. We verify this relation in \ref{fig:CovLevy} as well as the hypothesis in \ref{fig:CovT1T2}. Even though there is evidence that $\tau_k$ have algebraically decreasing correlations, since the condition \eqref{eq:Criterion} is respected, the time behavior of the covariance still holds. 
\begin{figure}[th!]
    \centering
    \includegraphics[width=\columnwidth]{Corr.pdf}
    \caption{{\bf Correlation functions of $1d$ L\'evy flights' $\tau_k$}:\\
    Upper: Computation of the covariance $Y=\text{Cov}[\tau_{n_1},\tau_{n_1+k}]/\text{Var}(\tau_{n_1})$ as a function of $k$. Errors bars are the $95\%$ confidence intervals.\\
    Lower: Summed covariance divided by $n_1^{2\alpha}$,  $Y\equiv \frac{1}{n_1^{2\alpha}}\text{Cov}\left[\sum_{k=0}^{n_1-1}\tau_k,\sum_{k=n_1}^{n_2-1}\tau_k \right]$ as a function of $n_2$ for different values of $n_1$ (in blue $n_1=10$, in orange $n_1=100$ and in green $n_1=1000$). \\
    The $\alpha$ values are $\alpha=1.3$ (left), $\alpha=1.5$ (middle) and $\alpha=1.7$ (right).}
    \label{fig:CovT1T2}
\end{figure}
We emphasize that in the case in which this criterion is not respected, the covariance between distinct sites visited has still a lower bound given by \eqref{eq:Covktimes}, as we show for the T-tree in \ref{fig:CovTtree}. Despite the noisiness of the data for  $\frac{1}{n_1^{2\mu}}\text{Cov}\left[\sum_{k=0}^{n_1-1}\tau_k,\sum_{k=n_1}^{n_2-1}\tau_k \right]$, we observe that the covariance does not respect \eqref{eq:Criterion}.
\begin{figure}[th!]
    \centering
    \includegraphics[width=\columnwidth]{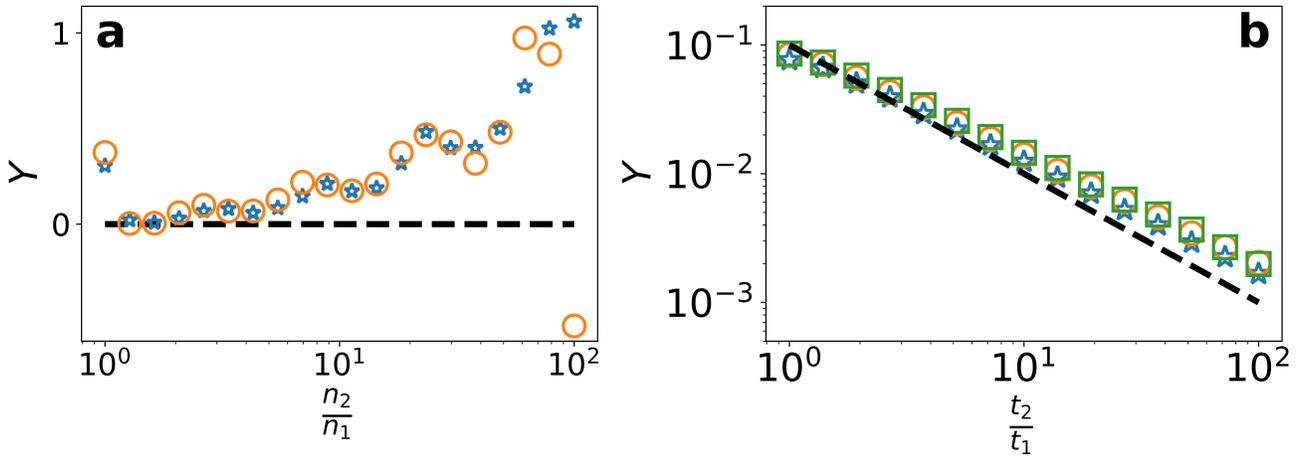}
    \caption{ {\bf Correlation functions of the T-tree:} {\bf a} $Y=\frac{1}{n_1^{2/\mu}}\text{Cov}\left[\sum_{k=1}^{n_1}\tau_k,\sum_{k=n_1+1}^{n_2}\tau_k \right]$ as a function of $n_2$ ($n_1=10$ in blue stars, $n_1=100$ in orange circles). {\bf b } $Y=\frac{\text{Cov}\left[N(t_1),N(t_2)\right]}{\mathbb{E}(N(t_1))\mathbb{E}(N(t_2))}$ compared to a function proportional to $\frac{t_1}{t_2}$ ($t_1=10$, $100$, and $1000$ represented by blue stars, orange circles and green squares, respectively).}
    \label{fig:CovTtree}
\end{figure}
This is why the two-time covariance of the number of distinct sites visited is not the same as that for L\'evy flights. However, it still gives a lower bound on the scaling of the covariance. Thus for any recurrent random walk, the covariance is long range. For lattice random walks \cite{jain1974further} and recently for $\alpha$-stable random walks \cite{cygan2021functional} that satisfy $\mu>3/2$, it has been shown that the process $N(t)-\mathbb{E}(N(t))$ converges to a Brownian motion of variance $\text{Var}(N(t))$ from which we deduce the covariance. 
Indeed, noting $\sigma=\lim\limits_{t\to \infty}\frac{\sqrt{\text{Var}(N(t))}}{\sqrt{t}}$ and $\left( B_t \right)_{t \geq 0}$ the brownian motion such that $\text{Cov}\left[B_{t_1},B_{t_2} \right]=\min \left(t_1,t_2 \right)$, we have that 
\begin{align}
    \text{Cov}\left[N(t_1),N(t_2) \right] &\approx \sigma^2\text{Cov}\left[B_{t_1},B_{t_2} \right]\nonumber  \\
    &=\sigma^2 \min \left(t_1,t_2 \right)
\end{align}
The case $1 \leq \mu \leq 3/2$ remains an open question.
\\
\subsection{Non-Markovian RWs}
\subsubsection{CTRW}
\begin{figure}
    \centering
    \includegraphics[width=0.8\columnwidth]{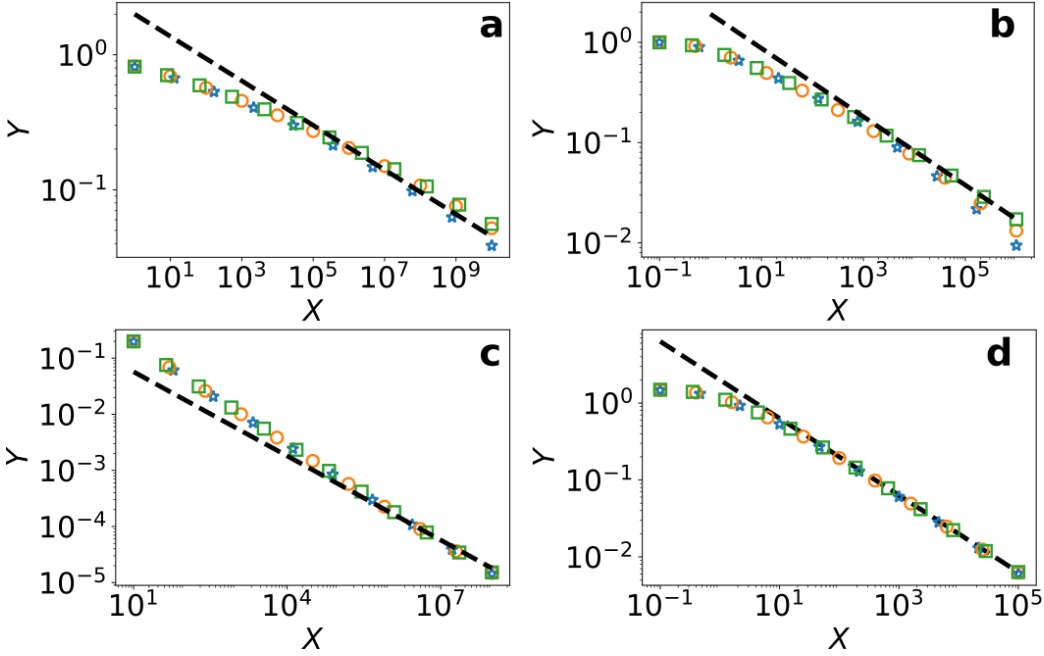}
    \caption{{\bf CTRW of recurrent nearest neighbour RWs:} {\bf a} on percolation clusters, $\beta=0.25$, {\bf b} on a Sierpinski gasket, $\beta=0.25$, {\bf c} on a $2d$ square lattice, $\beta=0.5$, {\bf d} on a a $3d$ hypercubic lattice, $\beta=0.5$. Each tail distribution $Y=\mathbb{P}(\tilde{\tau}_n >t)$ is compared to the predicted exponent of \eqref{eq:expCTRW} against $X=t$. For the marginal case {\bf c}, $Y=\mathbb{P}(\tilde{\tau}_n >t)/\ln t$.}
    \label{fig:CTRW}
\end{figure}

In this section, we show how the formalism of the main text, presented in the case of RWs in discrete time, can be extended to cover the important case of Continuous Time Random Walks (CTRWs). \\
For such walks, we notice that having the statistics of $\tau_n$ in the discrete setting (noted $F_n$) is enough to get the statistics $\tilde{\tau}_n$ in the continuous setting (noted $F^{\text{CTRW}}_n$), as adding the waiting times between jumps do not modify the statistics of the set of $n$ visited sites. By defining $\lambda_k$ as the time associated to the $k^{\rm th}$ displacement of the random walker:\\
%\begin{align}
%    \mathbb{P}(\tilde{\tau}_n=t)=\sum_{k=1}^{\infty} \mathbb{P}(\tau_n=k) \mathbb{P}(\lambda_1+\ldots+\lambda_k=t) \; .
%\end{align}
\begin{align}
    F^\text{CTRW}_n(t)=\sum_{k=1}^{\infty} F_n(k) \mathbb{P}(\lambda_1+\ldots+\lambda_k=t) \; .
\end{align}
We will be mainly interested in the case where  $\lambda_k's$ are i.i.d. with power law tails, $p(\lambda_k \geq t)\propto t^{-\beta}$ at large time, with $\beta<1$. They represent waiting times with infinite average. To simplify the calculations, we will assume that $\lambda_k$ are drawn from a L\'evy stable law of parameter $\beta$ such that $\lambda_1+\ldots+\lambda_k$ and $k^{1/\beta} \lambda_0$ have the same distribution. We compute the scaling behaviours of the distribution of $\tilde{\tau}_n$ in the following manner for the recurrent and marginally recurrent random walks:
\begin{align}
    \mathbb{P}(\tilde{\tau}_n \geq t)&\approx \sum_{k=1}^{t_n} \frac{A}{k^{1+\mu}}\mathbb{P}(k^{1/\beta} \lambda_0 \geq t)\nonumber \\
    &\approx \sum_{k=1}^{\min(t_n,t^\beta)} \frac{A}{k^{1+\mu}}\frac{Bk}{t^\beta} + \sum_{k=\min(t_n,t^\beta)}^{t_n} \frac{A}{k^{1+\mu}}
\end{align}
where we have been using that for L\'evy stable laws, the tail distribution is $\sim \frac{B}{t^{\beta}}$ for $t\gg 1$. We neglect the exponential and stretched exponential regimes occurring at $k>t_n$. We consider first the case, $t^\beta>t_n$: 
\begin{equation}
    F_n^\text{CTRW}(t)\propto -\frac{{\rm d}}{{\rm d}t} \left( \sum_{k=1}^{t_n} \frac{A}{k^{1+\mu}}\frac{Bk}{t^\beta} \right)
    \propto \left\{
    \begin{array}{ll}
         \frac{t_n^{1-\mu}}{t^{\beta+1}}\propto \frac{n^{1/\mu-1}}{t^{\beta+1}}\quad  \mbox{ if } \mu<1 \\
          \frac{\ln t_n}{t^{\beta+1}} \propto \frac{\ln n}{t^{\beta+1}} \quad \mbox{ if } \mu=1\;.
    \end{array}
\right.
\label{eq:expCTRW}
\end{equation}
while for $t^\beta<t_n$,
\begin{equation}
\label{eq:CTRWscaling}
    F_n^\text{CTRW}(t)\propto -\frac{{\rm d}}{{\rm d}t} \left( \sum_{k=1}^{t^\beta} \frac{A}{k^{1+\mu}}\frac{Bk}{t^\beta} + \sum_{k=t^\beta}^{t_n} \frac{A}{k^{1+\mu}} \right) \propto \left\{ \begin{array}{ll}
         \frac{1}{t^{\beta\mu+1}} \quad \mbox{ if } \mu<1 \\[2mm]
           \frac{\ln t}{t^{\beta+1}} \;\quad \mbox{ if $\mu=1$\;.}
    \end{array}
\right.
\end{equation}
 We check this regime in \ref{fig:CTRW} for different values of $\mu$ and $\beta$. The distribution $F_n^\text{CTRW}$ of $\tilde{\tau}_n$ is always algebraic, but the exponent changes at $t=t_n^{1/\beta}$: it is first algebraic of power $-1-\beta \mu$, and then of power $-\beta-1 $ as the distribution of the waiting times. It results from a competition between the algebraic waiting times and the algebraic distribution of $\tau_n$. Adding one additional waiting time has an algebraic cost. However, when $t^\beta$ is large compared to the cut-off time $t_n$, adding one waiting time has an exponential cost: thus, it is more likely to obtain large exit time because one waiting time is large than because $\tau_n$ is large. Thus, the exponential and stretched exponential regimes play no role in the algebraic law. This is why for transient walks, $\mu>1$, the law of $\tilde{\tau}_n$ is simply algebraic of parameter $-1-\beta$ as can be seen in \ref{fig:CTRW}: only a few values of $\tau_n$ matter. It is much more likely to obtain a large value of $\tilde{\tau}_n$ because one waiting time dominates than because $\tau_n$ is large.
We can thus generalize our result to CTRWs (in particular for $\beta<1$) by switching the exponent $1+\mu$ of Markovian RWs to $1+\beta\mu$, which translates into $1+\beta(1-\theta/\beta)$ when the algebraic decay is expressed in terms of the persistence exponent $\theta$.
\bigskip 
\subsubsection{$1d$ L\'evy flights with crossing }

In this section, we show how the formalism of the main text, presented in the case of Markovian RWs for which a renewal type equation between the propagator and the first-passage time density holds, can be extended to cover the important case of $1d$ Lévy flights that find sites with the so-called "crossing" convention (to be opposed with the "arrival" condition, used implicitly in the main text when referring to Lévy flights). While the renewal equation holds for Lévy flights with the arrival condition, it does not for Lévy flights with the crossing condition. Here we propose a direct analysis  of the $1d$ Lévy flights with the crossing condition. Importantly, we show that our results (and in particular the exponent of the algebraic decay) still apply in this case. \\

\begin{figure}
    \centering
    \includegraphics           [width= 0.6\columnwidth]{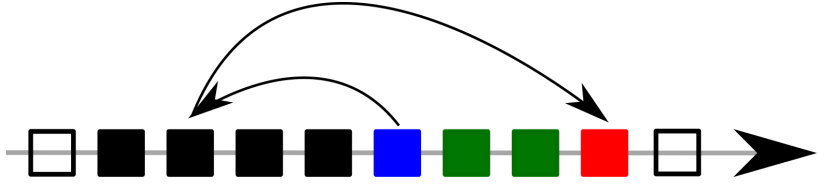}
    \caption{We consider the example where 5 sites have been visited (4 black squares and 1 blue where the RW starts). The L\'evy flight performs two jumps before arriving on the red site it never visited before. During the second flight, the walker flew above two other previously unvisited sites (in green). Thus, $\tau_5=2$, $\tau_6=0$, and $\tau_7=0$. $\tau_8$ is given by the first exit time from the interval of 8 squares starting from the red one.}
    \label{fig:LevyCross}
\end{figure}

\begin{figure}
    \centering
    \includegraphics[width=0.5\columnwidth]{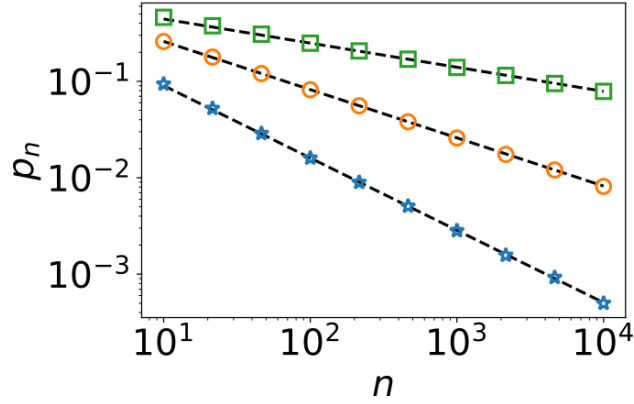}
    \caption{$p_n$ for $\alpha=0.5$, $\alpha=1$ and $\alpha=1.5$ (blue stars, orange circles, and green squares, respectively) compared to their predicted dependence on $n$, $n^{\alpha/2-1}$.}
    \label{fig:pnWalk}
\end{figure}

We consider a L\'evy flight, for which every step is drawn from a Zif's law $p(l)=\frac{1}{2\zeta(1+\alpha)|l|^{1+\alpha}}$, and contrary to what we presented in the main text Fig.~3{\bf a} of the main text, all sites the walker flies above are visited (the visited region contains no holes). Thus, the analysis on $\tau_n$ can be separated in two parts:
\begin{itemize}
    \item The walker is 'flying' when it visits the $n^{\rm th}$ site:  in this case, $\tau_n=0$.
    \item The walker stopped exactly at the last visited site $n$. Thus, the time distribution of exit times is given by the exit time distribution of a L\'evy flight from an interval of size $n$, starting at distance 1 from the border. We write this distribution as $\tilde{F}_n(t)$.
\end{itemize}
We illustrate this definition in \ref{fig:LevyCross}.
We define $p_n$ as the probability for the walk to stop exactly at the $n^{\rm th}$ visited site. Thus the distribution $F_n$ of $\tau_n$ can be described as 
\begin{align}
\label{distribLevyWalk}
F_n(t)=\delta(t)(1-p_n)+p_n \tilde{F}_n(t) \; .
\end{align}
This results in 
\begin{align}
\label{MomentLevyWalkReq}
\left \langle \tau_n^k  \right \rangle \sim p_n \int_0^{\infty}  t^k \tilde{F}_n(t) {\rm d}t  \; .
\end{align}
We use the result from \cite{levernier2018universal} for the distribution of the first hitting time of a target at distance $r=1$ in a line of length $R=n/2$ with a reflecting boundary condition. It is indeed equivalent to the first hitting time of one boundary of a finite domain of length $n$ starting at distance one from one border. From this, we get that  
\begin{align}
\label{MomentLevyWalkBorder}
\int_0^{\infty}  t^k \tilde{F}_n(t) {\rm d}t \sim  n^{k \alpha - \alpha/2  } \; .
\end{align}
Next, we derive the scaling with $n$ of the first moment of $\tau_n$ from the scaling with $t$ of $N(t)$. $N(t)$ is given by the difference between the maximum and the minimum of the position of the L\'evy flights after $t$ steps, which are know to scale as $t^{1/\alpha}$ \cite{bouchaud1990anomalous}. We note that for $\alpha<1$, the average of $N(t)$ is infinite, but the average of $N(t)^q$ for $q$ small enough is finite and given by $t^{q/\alpha}$. This is why in the following we replace $N(t)$ by $t^{1/\alpha}$ for any value of $\alpha$. By using the approximate relation 
\begin{align}
    \frac{{\rm d} N(t) }{{\rm d} t} \propto \frac{1}{\left\langle \tau_{ N(t) } \right\rangle}
\end{align}
we obtain that
\begin{align}
\left \langle \tau _n \right \rangle \sim
       n^{\alpha -1 } \sim p_n n^{\alpha/2  } \; .
\end{align}
It implies that for any value of $\alpha$,
\begin{align}
\label{pnDetermination}
p_n \sim n^{\alpha/2-1} \; ,
\end{align}
which we check on the figure \ref{fig:pnWalk}.
Combining \eqref{MomentLevyWalkReq}, \eqref{MomentLevyWalkBorder} and \eqref{pnDetermination}, we get 
\begin{align}
\left \langle \tau_n^k  \right \rangle \sim n^{k \alpha -1 }  \; . 
\end{align}
 Besides, using \eqref{distribLevyWalk} and \eqref{pnDetermination}, we have that $\tau_n$ follows a scaling behaviour in the limit $t,n \to \infty $, $t/n^\alpha$ fixed:
\begin{align}
F_n(t) \sim 
        \frac{1}{n^{1+\alpha}}\psi \left(t/n^\alpha \right) \; ,
\end{align}
where $\psi$ is algebraic at small $x=t/n^\alpha$, of exponent $-1-\theta=-1-1/2$ where $\theta$ is the persistence exponent \cite{levernier2018universal},
\begin{align}
    \psi(x)\sim x^{-1-1/2} \; .
    \label{eq:LevyFlightCross}
\end{align}
In particular, the formula given in Table 1 of the main text for the exponent of the algebraic decay at early times is valid in the case of this non-Markovian random walk. 

\subsubsection{L\'evy walks}
In this section, we extend the result of L\'evy flights with crossings to describe the L\'evy walk. From the result on L\'evy flights with crossings, we deduce the behaviour of $\tau_k$ for L\'evy walks, where the jumps are not instantaneous but have a duration equal to the length of the jump. For $\alpha>1$, the result is unchanged because the waiting times have finite mean, but for $\alpha<1$, by applying what we did in S5.C.{\it1} for CTRW to our situation, we obtain that the early-time regime is still algebraic but with a modified exponent (here, $\beta \to \alpha$ and $\mu \to 1/2$), 
\begin{align}
F_n(t) \sim \frac{1}{n^{2}}\phi \left(t/n \right) \; ,
\end{align}
where $\phi$ is algebraic at small $x=t/n$, of exponent $-1-\alpha/2$
\begin{align}
    \phi(x)\sim x^{-1-\alpha/2} \; .
    \label{eq:LevyWalk}
\end{align}
Once again, we find that the exponent of Table 1 is still valid for this other type of non-Markovian random walk, which strongly confirms the exponent $1+\beta(1- \theta/\beta)$ (for CTRW with infinite waiting times, $\beta<1$, and otherwise $\beta=1$).
\subsubsection{Algebraic decay for non-Markovian recurrent random walks}
Because of the applicability of our result to the non-Markovian cases of L\'evy flights with crossing and L\'evy walks for any value of $\alpha$ in $1d$, we test the exponent $2-\theta$ (for walks with waiting times of finite mean) of the algebraic decay of the distribution $F_n(\tau)$ in three other cases: the sub-diffusive Fractional Brownian Motion (characterized by the Hurst exponent $H$ such that $d_\text{w}=1/H>2$), the super-diffusive Fractional Brownian Motion ($d_\text{w}=1/H<2$) and the True Self Avoiding Random Walk \cite{barbier2020anomalous} ($d_\text{w}=3/2$).
We simulate the Fractional Brownian motion by using the module fbm \cite{fbmModule} of python based on Hosking’s method \cite{Hosking}. We discretize the line in intervals of size one, and consider that an interval has been visited when the RW enters it for the first time. $\tau_n$ is the time elapsed between visit of the $n^{\rm th}$ interval and the new $(n+1)^{\rm st}$ interval. For the True Self Avoiding Random Walk, we record the number of visits $C_i$ of site $i$. The probability to jump to the site on the right is given by $\exp(-C_{i+1})/(\exp(-C_{i+1})+\exp(-C_{i-1}))$, otherwise the RW jumps on the left.
We argue that $2-\theta$ is valid for a large range of non-Markovian RWs, despite the fact that the renewal equation we used to obtain this exponent is not valid anymore.\\

In Fig.~4{\bf g}, {\bf h} and {\bf i} of the main text, we show the algebraic decay of the probability $F_n(\tau)$ for the True Self Avoiding Walk and the Fractional Brownian Motion. Each is successfully compared to the algebraic decay $\tau^{-1-\mu}$ where $\mu=\frac{d_\text{f}}{d_\text{w}}$.
Based on the previous results for the Markovian and the non-Markovian RWs, we make the following prediction that for any recurrent random walks, the algebraic exponent governing the early time regime is given by 
\begin{align}
    F_n(\tau)  \propto 1/\tau^{1+\beta-\theta}
\end{align}
where $\theta$ is the persistent exponent, and $\beta$ is either $1$ or the exponent of the algebraic decay of the waiting time distribution of a jump $p(\lambda_k \geq t ) \propto t^{-\beta}$ if this exponent is smaller than $1$.\\
For discrete (in time) Markovian random walks, $\mu=\frac{d_\text{f}}{d_\text{w}}=1-\theta$ and $\beta=1$: we get back to the exponent $-1-\mu$ predicted in S3.A. For CTRW on Markovian RWs, $\theta=\beta(1-\mu)$ and we get back to \eqref{eq:CTRWscaling}. For $1d$ L\'evy flights with crossing, $\theta=1/2$ and $\beta=1$ while for L\'evy walks $\theta=\alpha/2$ and $\beta=\alpha$. We obtain again \eqref{eq:LevyFlightCross} and \eqref{eq:LevyWalk}. Finally, for True Self Avoiding Walks and Fractional Brownian motion $\theta=1-\mu$ and $\beta=1$ which gives again the exponent we confirm numerically in Fig.~4 of the main text.